\documentclass[12pt]{article}
\usepackage{epsfig}
\usepackage{a4,isolatin1}
\usepackage{amsmath,amsfonts,latexsym, amssymb}

\newtheorem{satz}{Theorem}[section]
\newtheorem{defi}[satz]{Definition}

\newtheorem{bem}[satz]{Remark}
\newtheorem{lemma}[satz]{Lemma}

\newtheorem{assumption}[satz]{Assumption}
\newtheorem{obdef}[satz]{Observation/Definition}
\newtheorem{conclusion}[satz]{Conclusion}
\newtheorem{ob}[satz]{Observation}

\newtheorem{conj}[satz]{Conjecture}
\newtheorem{prog}[satz]{Programme}

\newcommand{\mcal}{\mathcal}
\newcommand{\mbf}{\mathbf}

\newcommand{\tit}{\textit}

\newcommand{\R}{\mathbb{R}}

\newcommand{\Om}{\Omega}

\begin{document}
\thispagestyle{empty}
\begin{center}
\vspace*{1.0cm}

{\LARGE{\bf Let's call it Nonlocal Quantum Physics}}

\vskip 1.5cm

{\large {\bf Manfred Requardt}}\\email: requardt@theorie.physik.uni-goettingen.de 

\vskip 0.5 cm 

Institut f\"ur Theoretische Physik \\ 
Universit\"at G\"ottingen \\ 
Bunsenstrasse 9 \\ 
37073 G\"ottingen \quad Germany

\end{center}

\vspace{1 cm}

\begin{abstract}
  In the following we undertake to derive quantum theory as a
  stochastic low-energy and coarse-grained theory from a more
  primordial discrete and basically geometric theory living on the
  Planck scale and which (as we argue) possibly underlies also
  \tit{string theory}. We isolate the so-called \tit{ideal elements}
  which represent at the same time the cornerstones of the framework
  of ordinary quantum theory and show how and why they encode the
  \tit{non-local} aspects, being ubiquituous in the quantum realm, in
  a, on the surface, local way. We show that the quantum non-locality
  emerges in our approach as a natural consequence of the underlying
  \tit{two-storey} nature of space-time or the physical vacuum, that
  is, quantum theory turns out to be a residual effect of the
  geometric depth structure of space-time on the Planck scale. We
  indicate how the \tit{measurement problem} and the emergence of the
  \tit{macroscopic sub-regime} can be understood in this framework.
\end{abstract} \newpage
\setcounter{page}{1}
\section{Introduction}
In preceding work we have started to develop a radically discrete
mathematical and physical framework aimed at reconstructing, beginning
at the Planck scale and working ``bottom-up'' (so to speak), our
present day continuum physics and corresponding mathematics
(cf. \cite{1} to \cite{5}). Our main intention is it however to derive
both quantum theory and general relativity, i.e. gravitation (and in
the last consequence (semi)classical space-time), as emergent and low
energy \tit{effective theories} by a coarse graining process from a
more primordial discrete substratum.

While papers \cite{1} to \cite{4} deal mostly with the development of
the necessary mathematical and physical concepts and tools (typically
discrete protoforms of their continuum counterparts), some concrete
steps towards a realisation of the more ambitious latter goal were
taken in paper \cite{5} as far as the emergence and reconstruction of
a protoform of continuum space-time as, what we call, an \tit{order
  parameter manifold} is concerned. By this we mean an extended
coarse-grained \tit{superstructure} displaying a certain collective
order on a larger and smoother scale. As in the case of ordinary
\tit{order parameters} in, say, condensed matter physics, this
emergence of order is usually the result of a \tit{phase transition}
and is accompanied by a shrinking of microscopic phase space being
occupied by the system. Whereas this programme is far from
being completed, its core result (or rather: hypothesis, as not every
step in the corresponding analysis is, up to now, rigorously proved)
can be summarized as follows.

The physical vacuum or (semiclassical) space-time has to be considered
on a certain level of resolution as a two-story structure. It consists
of a relatively smooth ``surface layer'' formed by an intricate web of
overlaping \tit{lumps} (the \tit{physical points}) and representing the
quasi-continuous medium we experience as ordinary space-time. Beneath
this surface there exists a more irregular and wildly fluctuating
``underworld'' of a distinctly discrete and stochastic nature
(stochastic compared to our ordinary level of resolution; at the very
bottom the underlying dynamics may well be deterministic!). Its
perhaps most characteristic feature is a peculiar \tit{non-local}
dynamical behavior observed in \cite{5} and further analyzed below as
it plays a decisive role in the understanding of quantum theory. Each
of these two stories has its own physical or dynamical \tit{distance
  function} or \tit{metric} and, typically, there will exist a certain amount of
direct interactions or exchange of information in this mentioned
underground between regions (or lumps) lying a certain distance apart
with respect to the distance concept holding sway in the coarse
grained surface structure (i.e. our classical space-time).

Our whole approach is technically based on what we call a
\tit{cellular network}, dubbed $QX$ for short (``quantum space''), as
the most primordial substratum in our framework. This network is
assumed to have been evolved in the distant past (``\tit{big bang}'')
from a certain chaotic \tit{initial phase} denoted by $QX_0$ (and
which is characterized, among other things, by the complete absence of
stable patterns) through a regime of geometric change (called a
\tit{geometric phase transition zone}) into a new phase $QX/ST$. This
latter phase represents the above described two story
\tit{superstructure}, i.e. the underlying primordial network
superposed by a coarser network consisting of a web of lumps, that is,
certain subgraphs with a particularly dense internal connectivity
among the respective nodes (and playing the role of the ``\tit{physical
points}''). These physical points are considered to be the constituents
of the relatively smooth surface structure, $ST$. This new phase,
$QX/ST$, is the epoch our universe is roaming in since the moment when
space-time emerged from this mentioned underground as an approximately separate
entity.

It is a peculiar feature of this kind of geometric phase transition
(described in much more detail in \cite{5}) that it equips the
mentioned surface- or superstructure of lumps with a so-called
\tit{Nahwirkungsprinzip}, while on a finer level of resolution there
remain a lot of additional non-local interactions among distant lumps of,
however, a more subtle nature. We will argue in the following that
this almost hidden non-local web of exchange of information, which
arises quite naturally in our approach, plays a decisive role in the
formation of quantum theory as an effective continuum theory
incorporating certain non-local gross features of the depth structure
of space-time (about such a possibility was already speculated in
\cite{Rosen} --  and possibly also elsewhere -- as a way out of the so-called EPR-paradoxon).

It goes without saying that, given the complexity of the task and the
long and entangled history of the subject, such things cannot simply
be proved in a rigorous sense of the word as there does not even exist
a universally accepted framework from which to start, not even in
ordinary (non)orthodox quantum theory. By necessity our approach has
to be, at least in this preparatory stage, speculative to some extent
and has to be based on a more or less loose (or strong, depending on
the point of view) web of arguments mixed with a certain amount of
``educated guesswork''.

Furthermore, as the interpretation or epistemology of quantum theory
has such a long and involved history of its own we have to refrain
from recapitulating too much of this nightmarish and mind boggling
subject.  That is, lack of space prevents us from giving full credit
to many researchers in this field. This would afford a full monograph
and would give the paper a perhaps too ``philosophical'' touch. We
therefore concentrate in the rest of this introductory section on
mentioning and quoting in loose order those approaches and points of
view which appear to be similar (at least to some extent) in spirit to
our own working philosophy and make some comments and anotations. In
the next section we discuss two approaches in slightly more detail as
they are related a little bit closer to our own one in several
technical respects.

We begin this brief historical part with two general remarks by von
Weizs\"acker (\cite{33}, similar ideas were also entertained by
Wheeler, see e.g. \cite{Wheeler}) which strike the key of our paper.  
\begin{quote}
{\small \ldots space-time is not the background but a surface aspect of
reality\ldots It is extremely improbable that this reality
(i.e. quantum reality) will be describable as consisting of events
which are localized in space and time.

The translocal phase relations are ``surplus information'' not lack of
information. Quantum theory knows more, not less, than local classical
physics.}
\end{quote}

A well known critic of the orthodox Copenhagen interpretation was
Einstein. Here are some illuminating utterances which are in our view
very much to the point. More about his original scientific attitude
can be found in the beautiful essay of Stachel in \cite{6} or \cite{7}
\begin{quote}
{\small It is\ldots to be expected that behind quantum mechanics there
  lies a lawfulness and a description that refer to the individual
  system. That it is not attainable within the bounds or concepts
  taken from classical mechanics is clear.

I do not at all doubt that the contemporary quantum theory (more
exactly ``quantum mechanics'') is the most complete theory compatible
with experience, as long as one bases the description on the concepts
of material point and potential energy as fundamental concepts.}
\end{quote}
(The latter remark is taken from \cite{7}). We want to stress the
fundamental importance of the underlying insight being conveyed in
these remarks. It is a crucial observation that quantum mechanics (as
we know it) happens to be just the description of the ``quantum
world'' if one starts from the core concepts of classical mechanics
like e.g.  position, momentum etc. Following this line one may get a
very specific and biased class of observables while excluding other
possible elementary concepts. This contextual and historical
dependence of theories, frameworks and whole working philosophies is
frequently overlooked or, at least, not sufficiently appreciated. The
problems we still have with quantum theory may just result from a too
selective choice we have made in the past.

In other words, while quantum theory has made the first steps away
from the mechanistic \tit{particle picture}, it has still retained
many of its conceptual ingredients and has molded it into the hybrid
of the so-called \tit{wave-particle duality}. A similar point of view
was hold by e.g.  Schr\"odinger (see his beautiful and extensive
biography, \cite{9}).

A further point worth of mentioning is Einsteins open attitude
towards the \tit{discrete} and the \tit{continuum}. Many quotations
can again be found in (\cite{6} p.27ff). Another source (mentioned
already in \cite{5}) is his commentary in \cite{8} on the contribution
of Menger (\tit{geometry of lumps} and \tit{statistical metrical space}).

Haeretic views were also hold by Dirac, a fact which is perhaps not so
widely known (cf. e.g. his biography, \cite{11} p.201ff). He in fact
tried for many years to revive a modern \tit{aether concept} as a
common receptacle for all the physical processes (see also
\cite{12}). Similar ideas were uttered by Bell (\cite{13}) or T.D.Lee
(\cite{14}), to mention a few.

The arguments for the absence of such an underlying substrate are in
our view far from being convincing and are rather typical for what is
called a paradigm, i.e. an adopted working philosophy which, when
accepted, tends to become very rigid and constellates and frames our
whole attitude towards the occurring phenomena and their codification
in the form of theoretical concepts. We think that this paradigm is
responsible to some extent for the interpretational difficulties and
seemingly paradoxical language of orthodox quantum mechanics. We are
convinced that quantum mechanics would become considerably less
paradoxical if we were prepared to realize the ubiquituous
interference phenomena (which are in fact the pivotal point of quantum
theory) and the complex structure as part of the quantum information
conveyed by extended excitation patterns roaming this largely hidden
``underground''.

One of the consequences of taking the possibility of such a hidden and
subtly organized substructure not taking into account and regarding,
instead of that, space-time as the primordial receptacle is the
attitude to consider e.g. \tit{wave functions} and their seeming
breakdown as mere subjective artifacts. This becomes particularly
apparent in the usual discussion of the \tit{double slit
  experiment}. In our view it is difficult to deny that there is
``something'' passing ``through'' both slits in the undisturbed situation
in each individual experiment. But this ``something'' cannot be so
easily detected as long as the intricate substructure of space-time is
not realized.

It is no question that the old (mechanical) aether concept is almost
empty, but its emptyness resulted from another even older paradigmatic
preoccupation of that time, i.e. the fiction of an empty, a priori and
independently existing geometric \tit{background space} which is then
permeated by some medium called aether. One should rather regard
(background) space-time as only a part of an underlying more complex
substrate in the way we have indicated above.

This whole bundle of problems and ideas belongs in fact to a much
wider topic, which originated already with Leibniz and Mach (see e.g.
the lively debate in \cite{15}) and has been lucidly clarified by
Einstein. It is the almost universal topic of the role of so-called
\tit{ideal elements} in scientific theories. As their role in quantum
theory will be a central theme of our paper we give it a closer
inspection in the next section and close the introduction with some
(as we think) deep remarks and reservations uttered by Scrooge in
\cite{16} to which we fully subscribe.
\begin{quote}
{\small Wave functions are real for the same reason that quarks and
  symmetries are\ldots Any system is in a definite state \tit{whether
    any humans are observing it or not}; the state is not described by
  a position or a momentum but by a wave function.

It seems to me that none of this forces us to stop thinking of the
wave functions as real, it just behaves in ways that we are not used
to, including instantaneous changes.

(Weinbergs own utterance on p.143): The positivist concentration on
observables like particle positions and momenta has stood in the way
of a realist interpretation of quantum mechanics in which the wave
function is the representation of physical reality.}
\end{quote}

The following sections, 2 to 4, are of a preparatory nature, that is,
they provide the necessary background, motivations and concepts, thus
paving the ground for the central sections, 5 and 6, of the paper in
which quantum theory is derived as a low-energy effective theory of a
more primordial theory, living on the Planck scale. In a short aside
we speculate about the possibility that this fundamental theory may
also underly \tit{string theory} (near the end of section 5). In
section 7 we briefly indicate in what directions our appoach has to be
further developed in order to address the so-called \tit{measurement
  problem} and/or the emergence of \tit{macroscopic behavior}.

\section{The Description of our own Working Philosophy}

In this section we want to discuss the pieces of our own working
philosophy in more detail and relate it to two other approaches
presented in the more recent past, which seem to be developed in a
similar spirit. The one of the mentioned approaches is expounded in
two longer papers by Smolin (\cite{17},\cite{18}) and stems from the
period before he embarked on the \tit{loop quantum gravity program}.
The other is the work of 't Hooft about a presumed \tit{cellular
  automaton} substrate underlying quantum theory (see \cite{27} to
\cite{29}). While these two frameworks differ in several respects from
each other, each of them shares, on the other side, a bundle of ideas
with a certain particular strand in our own approach.
\subsection{A short Review of Smolin's Ideas, the Role of Ideal
  Elements and our own Point of View}
In this subsection we concentrate mainly on the epistemological and
foundational aspects of Smolin's work as they are of particular
importance for the understanding of the (sub)structure of quantum
theory. The more technical aspects and the concrete implementation are
postponed to the following sections. The greater part of the
epistemological ideas can be found in \cite{17}.

Central in this respect is the notion and role of \tit{ideal elements}
in physical model theories. Smolin argues that practically all our
theories contain - by necessity - so-called \tit{ideal elements} as
they typically deal only with a portion of our universe. He describes
them as absolute or background structures which are not themselves
determined by solving any dynamical equations or, put differently,
elements of the mathematical structure whose interpretation requires
the existence of things outside of the dynamical system described by
the theory. He argues that both \tit{Mach's Princicple} and the
\tit{quantum mechanical measurement problem} are cases in point, both
of which are crucial parts of the even greater problem of constructing
a sensible quantum cosmology. To give another but related definition,
we can say with Einstein that something that acts but is not acted
upon serves as an ideal element like e.g. \tit{inertial systems} in
special relativity (cf. \cite{19}).

The reason why this theme is so carefully discussed by Smolin is the
necessity to formulate a theory of quantum gravity that does not
depend on an absolute background space. This is the place where the
\tit{Machian Philosophy} comes into the play in form of a sceptical
attitude towards the existence of space-time as a metaphysical and a
priori substratum. The reader who is interested in the actual content
of this (a little bit poetic) principle should consult \cite{15} or
e.g. \cite{20}. Another illuminating characterisation, stemming from
Westpfahl (\cite{22}), is quoted in \cite{21}.
\begin{quote}
{\small \ldots all tendencies which try to reduce all the phenomena
  which cannot be described by laws of nature (viz. field equations)
  to cosmological causes}
\end{quote}
As our primary concern in this paper is the creation of an underlying
more primordial theory which contains ordinary quantum theory as an
effective and derived stochastic theory, we will henceforth
concentrate more on the role of ideal elements in quantum theory.
Perhaps a little bit surprisingly, we will show later that in a
certain way such a Machian strategy, i.e. explaining seemingly local
features of a theory by a \tit{nonlocal} influence of the (distant)
environment, will also work in quantum theory.

Now, what are the ideal elements in quantum theory? Deviating slightly
from the analysis of Smolin we concentrate at the moment not so much
on the infameous measurement problem but on another (in our view)
central structural element of quantum theory, i.e. the
\tit{superposition principle} together with the genuine \tit{complex
  structure} of the theory.
\begin{conj}[Ideal Elements in Quantum Theory]\hfill
\begin{enumerate}
\item In a similar sense as Smolin did, we conjecture that the complex Hilbert
  space structure of ordinary quantum theory (and in particular the
  superposition principle) are playing the role of ideal elements in
  quantum theory. They encode in a {\em local way} a {\em
    nonlocal} stochastic interaction between the lumps of the surface
  structure $ST$.
\item Furthermore we claim that a good deal of the observed local
  quantum fluctuations and randomness has its origin in the fact that
  ordinary quantum theory is, by necessity, that is, by its very
  definition, a theory of only a small portion of the universe, with
  this portion being open to permanent {\em nonlocal} interaction with
  the (distant) regions of the quantum environment.
\item Put differently, as in the Machian concept of inertia, we assume
  that the mentioned ideal structural elements of quantum theory
  encode in a, on the surface, local way nonlocal effects which
  originate on a more primordiallevel and which make the {\em local
    version of the theory}, formulated in macroscopic space-time, a
  {\em stochastic} one. The deeper reason for this is that this local
  formulation describes (while being unaware of it) a, in some sense, {\em open system}.
\end{enumerate}
\end{conj}

The last remark leads to another central theme of a \tit{realistic
  approach} towards quantum theory, viz. the nature and origin of
statistics in the quantum realm. Smolin remarks in \cite{18} by
referring to various recent observations in quantum cosmology
(e.g. \cite{23}) that there seems to be no local coordinate invariant
distinction between \tit{real} statistical fluctuations (in the
``ordinary'' sense) and \tit{virtual} quantum fluctuations. The lesson
we learned from Einstein is then the following.
\begin{ob}
If the distinction of two phenomena depends on the system of reference
then these, superficially different, phenomena are actually of the
same kin.
\end{ob}
Consequently our program can be described as follows.
\begin{prog}
Find an underlying more primordial model theory in which {\em virtual
quantum fluctuations}  are {\em ordinary statistical fluctuations}.
\end{prog}

This is now the point where the various strategies bifurcate from each
other. In \cite{18} Smolin chose to further develop the Nelson
progamme of stochastic (quantum) mechanics, which is based on
the picture of a particle moving in a quasi-brownian environment with,
however, a quite peculiar \tit{diffusion behavior} not observed in the
classical regime (as to this programme see e.g. \cite{24} or
\cite{25}). Smolin then embeds the system in a background
gravitational field and argues that this strategy shows that and how
the Hilbert space framework has to be transcended. At the end of the
paper he discusses certain models dealing with nonlocal hidden
variables.

We should say that this was, at least to some extent, a strategy we
also persued in earlier times. In \cite{26} we argued for example that
the Nelson approach has to be developed further in the direction of
\tit{nonmarkovian mechanics} and made some tentative steps along these
lines. Our main argument was that the fluctuating but \tit{passive}
background in the Nelson-approach (viz. an ideal element par
excellence) has to become dynamical due to the (back)reaction of the
randomly moving quantum particle, which, in our view, can no longer be
neglected in the quantum regime. If one now averages over this
dynamical background one would get an evolution equation for the
particle itself which contains now a \tit{memory kernel} (as the
averaging process will collect retardation effects), i.e. which becomes
\tit{nonmarkovian}. The technical implementation of this programme,
however, turned out to be extraordinarily ambitious and we hesitated
at that time to push the work beyond the preprint status.

What all these and related approaches are having in common is that
they stick to a particle concept as fundamental building block, i.e.
it is assumed that there exists a discrete entity which moves in a
fluctuating background. This is, however, exactly the concept we
choose to abandon in our more recent approach in which quantum objects
are, on the contrary, assumed to be extended excitation patterns
roaming our two story network environment but carrying certain
discrete particle properties which are observed in suitable
measurement set-ups. On the other hand, it may well be that the older
ansatz turns out to be a certain useful approximation to this more
complex picture.
\subsection{'t Hooft's Framework plus some Comments}

The starting point for 't Hooft is a little bit different from the
preceding approach. His emphasis lies on the discreteness of physics
at the Planck scale and is thus related to the other strand of ideas
on which our own framework is based. It is his aim to derive quantum
behavior on a large scale (i.e. also by a kind of coarse graining)
from suitable deterministic \tit{cellular automaton laws} holding sway
on a more primordial scale (see \cite{27} to \cite{29}). There exist
earlier related ideas scattered in the literature which were inspired
by the concept of cellular automata (e.g. \cite{30} or \cite{Zuse}). In his most
recent contribution (\cite{29}) he argues that certain versions of
hidden variable theories must be revived in the face of problems in
quantum gravity and that space, time and matter all have to be
\tit{discrete} at bottom. 

Another point he rightly emphasizes is that (obviously) the primordial
degrees of freedom are \tit{not} describing electrons or any other
particles, but microscopic variables at scales comparable to the
Planck scale. This is exactly in line with what we said above and with
our own framework. We will readdress this particular aspect in the next
subsection under the catchword of \tit{the problem of scales}.

Of particular relevance for our enterprise is the following somewhat
related argument against the many critics of such deterministic
approaches in the quantum realm. `t Hofft reasons that quantum theory
provides a completely adequate framework on its natural scale of
resolution and that there is no chance to replace it on this scale by
some classical or deterministic model theory. But it may well be that
a model theory being deterministic on the Planck scale generates the
statistical quantum laws via coarse graining on their natural scale,
thus invalidating the consequences of, say, the \tit{Bell inequalities}.

While we are very sympathetic with this programme we would like to
comment on some differences as compared to our own framework. The main
difference, we think, is that we do not base our analysis on a
rigid a priori fixed lattice structure but, instead of that, regard
the geometric wiring diagram underlying the cellular network as a full
fledged dynamical system of its own which interacts with the node
states (which, on the other side, are the only variables in a cellular
automaton). Thus geometry, dimension,metrical properties, near- and
far-order all become dynamical \tit{collective quantities} which are
assumed to \tit{coevolve}. The underlying philosphy is of course that
ultimately both quantum theory \tit{and} gravity emerge as two
seemingly different but in fact related large scale aspects of one
and the same underlying theory.

Another important aspect of our programme is that we will show in the
following that quantum theory encodes in a at first glance local way on its
own natural scale hidden \tit{nonlocal long-range} interactions among the really
microscopic degrees of freedom, living on a more primordial level.
\subsection{The Problem of Scales}

It is frequently argued that the attempt to relate e.g. quantum
physics on its presently accessible middle energy scales to some
underlying and largely hidden primordial theory, living on, say, the
Planck scale, is virtually impossible due to the huge difference in
orders of magnitude between the two regimes. This is called the
\tit{problem of scales}. There is certainly more than a grain of truth
in this criticism but we think one can turn this seeming difficulty
into an advantage by pursuing the following strategy.

As in the physics of the \tit{critical point} (in, say, statistical
mechanics or lattice quantum field theory), any continuum theory which
is nontrivial, that is, which has correlations and patterns extending
over non-zero scales, must necessarily be in a \tit{critical} or at
least \tit{near-critical} state on the microscopic scale, as all
finite length scales will shrink to zero in the \tit{continuum limit}.
That is, it must have very long range correlations on that scale,
which is typically only the case near or at the critical point.

By the same token, to a given continuum theory will belong a whole
\tit{universality class} of microscopic theories which lead to the
same macroscopic consequences. Applied to e.g. general relativity such
a point of view is expressed in \cite{31}. A perhaps even more radical
opinion is expressed in \cite{32} (there are in fact quite a few other
interesting ideas to be found in this book), running under the
catchword \tit{random dynamics}. The central hypothesis is that the
structure of the theories on the low- or middle-energy side of the
energy spectrum is to a large extent independent of the form
of the hypothesized fundamental theories on the ultra-high energy side
and that the structure of the former ones is rather a consequence of
the way how the coarse graining is performed.

This point of view is partly corroborated by the observations we make
below when we attempt to derive quantum theory from our network model.
On the other side this does not mean (at least in our view) that a
particular fundamental theory does not exist or that we shall be
unable to discriminate between different model theories in the future
(note that a similar standpoint could have been adopted with respect
to quantum theory as the underlying theory of classical mechanics).
The correct conclusion to draw is that it is not reasonable (in the
beginning) to concentrate too much on certain (possibly wrong or
unimportant) microscopic details but better have the gross features
right. That is, the real task may rather consist of extracting the
possibly few crucial characteristics which the primordial theory must
contain and which, in the end, survive the coarse graining limit.
\section{A brief R\'esum\'e of the Properties of the Two-Story Network
  Substratum}
\subsection{Some General Remarks}
As the technical details and underlying working philosophy can to a
large part be found in refs. \cite{1} to \cite{5} with special
emphasis on \cite{5}, we will be very brief. We emulate the underlying
substratum of our world, or, more specifically, of our space-time
(quantum) vacuum (containing however in addition all the existing
quantum and macro objects as extended excitation patterns!) by what we
call a cellular network.

This discrete structure consists of elementary \tit{nodes}, $n_i$,
which interact (or exchange information) with each other via
\tit{bonds}, $b_{ik}$, playing the role of (in this context) not
further reducible (abstract) elements. The possible internal structure
of the nodes (modules) or bonds (interaction chanels) is emulated by
discrete internal state spaces carried by the nodes/bonds. The node
set is assumed to be large but finite or countable. The bond $b_{ik}$
is assumed to connect the nodes $n_i,n_k$. The internal states of the
nodes/bonds are denoted by $s_i$, $J_{ik}$ respectively. As our
philosophy is, to generate complex behavior out of simple models we,
typically, make simple choices for them, one being e.g.
\begin{equation} s_i \in q\cdot \mathbb{Z}\quad,\quad J_{ik}\in
  \{-1,0,+1\} \end{equation}
with $q$ an elementary quantum of information.

As in our approach the bond states are dynamical degrees of freedom
which, a fortiori, can be switched off or on, the \tit{wiring}, that
is the pure \tit{geometry} of the network is also an emergent, dynamical
property and is \tit{not} given in advance. Consequently the nodes,
bonds are typically not ordered in a more or less regular array, a
lattice say, with a fixed nea-/far-order. This implies that
\tit{geometry} will become to some extent a \tit{relational} (Machian)
concept and is no longer an \tit{ideal element} (cf. the discussion in
sect. 2).

On the other side, as in cellular automata, the node and bond states
are updated (for convenience) in discrete clock time steps,
$t=z\cdot\tau$, $z\in\mathbb{Z}$ and $\tau$ being an elementary clock
time interval. This updating is given by some \tit{local} dynamical
law (examples given below). In this context \tit{local} means that the
node/bond states are changed at each clock time step according to a
prescription with input the overall state of a certain neighborhood
(in some topology) of the node/bond under discussion.  We want however
to emphasize that $t$ is \tit{not} to be confounded with some
\tit{physical time}, which, for its part, is also considered to be an
emergent coarse grained quantity. The well known \tit{problem of time}
is, for the time being, not treated in detail in the following, as it
is a big problem of its own, needing a careful and separate analysis
of its own (see however \cite{time} or \cite{Butter}). That is, at the
moment the above clock time is neither considered to be dynamical nor
observer dependent. There is however a brief discussion of the
presumed emergence of a new primordial time scale which sets the scale
for the regime where quantum fluctuations hold sway (see below).

In \cite{5} we gave examples of local dynamical laws which, we
presume, are capable of encoding the kind of geometric unfolding we
are expecting. An important ingredient is what we call a
\tit{hysteresis dynamics}, that is, the bonds, or more properly the
interactions $J_{ik}$, are switched off under appropriate conditions
of the local network environment. An example of such a local law is
the following:
\begin{defi}[Example of a Local Law]\label{law}
At each clock time step a certain {\em quantum} $q$
is exchanged between, say, the nodes $n_i$, $n_k$, connected by the
bond $b_{ik}$ such that 
\begin{equation} s_i(t+\tau)-s_i(t)=q\cdot\sum_k
  J_{ki}(t)\end{equation}
(i.e. if $J_{ki}=+1 $ a quantum $q$ flows from $n_k$ to $n_i$ etc.)\\
The second part of the law describes the {\em back reaction} on the bonds
(and is, typically, more subtle). This is the place where the
so-called {\em hysteresis interval} enters the stage. We assume the
existence of two {\em critical parameters}
$0\leq\lambda_1\leq\lambda_2$ with:
\begin{equation} J_{ik}(t+\tau)=0\quad\mbox{if}\quad
  |s_i(t)-s_k(t)|=:|s_{ik}(t)|>\lambda_2\end{equation}
\begin{equation} J_{ik}(t+\tau)=\pm1\quad\mbox{if}\quad 0<\pm
  s_{ik}(t)<\lambda_1\end{equation}
with the special proviso that
\begin{equation} J_{ik}(t+\tau)=J_{ik}(t)\quad\mbox{if}\quad s_{ik}(t)=0
\end{equation}
On the other side
\begin{equation} J_{ik}(t+\tau)= \left\{\begin{array}{ll} 
\pm1 & \quad J_{ik}(t)\neq 0 \\
0    & \quad J_{ik}(t)=0
\end{array} \right. \quad\mbox{if}\quad
\lambda_1\leq\pm
  s_{ik}(t)\leq\lambda_2 
\end{equation}
In other words, bonds are switched off if local spatial charge
fluctuations are too large, switched on again if they are too
small, their orientation following the sign of local charge
differences, or remain inactive.
\end{defi}
Remark: Another interesting law arises if one exchanges the role of
$\lambda_1$ and $\lambda_2$ in the above law, that is, bonds are
switched off if the local node fluctuations are too small and are
switched on again if they exceed $\lambda_2$.  We emulated all these
laws on a computer and studied a lot of network properties. The latter
law has the peculiar feature that it turned out to have very short
\tit{transients} in the simulations, i.e. it reaches an
\tit{attractor} in a very short clock time. Furthermore this
attractor turned out to be very regular, that is, it had a very short
period of typically six, the whole network returned in a previous
state after only six clock time steps, which is quite remarkable,
given the seeming complexity of the evolution and the huge phase space
(\cite{Nowotny}).\\[0.5cm]
Some characteristic features of these class of laws are the following.
\begin{ob}[Gauge Invariance]\hfill
\begin{enumerate}
\item The dynamics depends only on the local {\em charge differences},
  $s_i-s_k$ and nowhere on the absolute values $s_i$ itself, i.e. it
  is to some extent {\em relational}.
\item The {\em total charge}, $Q:=\sum_{n_i}s_i$, is conserved under
  the evolution. One could e.g. choose a boundary condition like
  $Q=0$, which may be considered as a kind of {\em gauge fixing}. 
\end{enumerate}
\end{ob}

The following point we consider to be of central importance,
irrepective of the concrete network law under discussion. 
\begin{ob}\label{geomatt} We expect that really interesting fundamental laws display
  the following generic patterns. They typically consist of more or less two
  parts, encoding the interaction of two primordial substructures,
  described a little bit sloppily by the catchwords {\em geometry} and
  {\em matter}.
\begin{enumerate}
\item {\em geometry} acting on {\em matter}
\item {\em matter} acting on {\em geometry}
\end{enumerate}
Usually the first part of the dynamical law seems to be relatively
simple and transparent, while the second part is typically much more
involved.
\end{ob}
Remark: Note that these criteria are fulfilled by our above example,
where the first part is more or less a \tit{conservation law}. The
geometric structure is the wiring of the network, i.e. the global bond
state. A classical case in point is \tit{general relativity}, where
the first part consists of the \tit{geodesic motion} of matter, the
latter part of the \tit{Einstein equations}. We will later show that
even quantum theory is already of this type if understood or looked
upon in a certain (new?) way.\vspace{0.5cm}

In \cite{5} we chose to concentrate on the geometric structure of the
network, thus neglecting most of the details of the microscopic
network state and the dynamics. In the corresponding reduced graphical
representation a bond, $b_{ik}$ was drawn between the nodes $n_i,n_k$
in the time-dependent graph, $G(V,E(t))$, iff the bond-state,
$J_{ik}(t)$, was different from zero ($V,E$ the set of nodes, edges
(bonds) respectively). In a next step these graphs were considered as
members of a certain probability space, $\mcal{G}(n,p)$, of
\tit{random graphs} over $n$ nodes and with the \tit{edge probability}
$0\leq p\leq 1$.
\begin{bem}One can as well choose a slightly higher resolution by
  keeping trace of the sign of the bond-interaction, that is
  $J_{ik}=+/-1$, and identify $J_{ik}=+1$ with a {\em directed bond},
  $d_{ik}$, pointing from $n_i$ to $n_k$ or vice versa for
  $J_{ik}=-1$. This would lead to a so-called {\em directed graph}.
\end{bem}
In \cite{5} we were particularly interested in certain subgraphs of a
typical random graph taken from $\mcal{G}(n,p)$, their size, number,
degree of overlap and entanglement. These particular subgraphs are
called \tit{cliques} in graph theory and are (in a technical sense)
maximal complete subgraphs or subsimplices, that is, all the pairs of
nodes belonging to a clique are connected by a bond and the cliques
are the \tit{maximal} elements in the respective chains of
subsimplices (ordered by inclusion). In more physical terms we also
called them \tit{lumps} or \tit{physical points}.

For later purposes we note that a graph carries a \tit{natural
  metric}:
\begin{defi}The (natural) distance, $d(n_i,n_k)$, between two nodes, $n_i,n_k$,
  is the length of a minimal path (a {\em geodesic}) connecting them,
  its length given by the number of bonds of the path. This distance
  defines a metric on G ($d:=\infty$ if the nodes are lying in
  different components).
\end{defi}
Remark: There are other interesting notions of distance one can
envisage on a graph. One is studied in \cite{3} and is related to
similar concepts in \tit{non-commutative geometry}. Another is
discussed at the end of \cite{17}.
\subsection{The Web of Lumps}
We argued in \cite{5} that what we experience as (quasi)classical
space-time and (on a higher resolution) as \tit{quantum vacuum},
consists roughly of two or rather three regimes. At the very bottom we
have the level of the primordial network with its corresponding
primordial length- and (clock) time scales, correlation lengths/times
etc. On the next level we have the web of lumps or physical points,
i.e. the web of overlapping cliques. this level defines a new group of
corresponding (length) scales, as it is usually the case if a new
phase emerges. We conjecture that these emerging scales are the
infameous \tit{Planck-units}, e.g. $l_P,t_P$ etc. On the macroscopic
surface level, which is the regime directly accessible to us, the
internal structure of the physical points is no longer visible, we
observe a (quasi) continuum as \tit{background space} which, on a
slightly finer scale, is roamed by \tit{quantum fluctuations},
representing the \tit{residual} low-energy effects of what is
happening on the deeper levels.

In \cite{5} we made a relatively detailed analysis of this web of
lumps within the framework of \tit{random graphs}. We calculated the
typical size of these lumps, their number, mutual overlap, expected
size of the infinitesimal neighborhood of a typical lump etc. On the
other side, some problems remained open for which we have, at the
moment, only partial answers (which is however no wonder, given the
enormous complexity of the behavior of the undrlying network). Note
that in the random graph approach we concentrated solely on the wiring
diagram of the network and studied its properties in a purely
statistical way. It became apparent that in order to follow its
dynamical evolution in more detail, something like a non-equilibrium
statitical mechanics for such systems is called for. Furthermore, the
pure random graph picture is possibly (or rather: probably) not
sufficient to explain every aspect of the \tit{unfolding process} of
the network towards the expected new phase, $QX/ST$. This, however,
has to be expected since we have learned that the unfolding towards a
level both of higher order and diversity may need some
\tit{fine-tuning} and is not the expected to be the ordinary situation
(a catchword being ``complexity at the edge of chaos'';
see\cite{Waldrop} to \cite{Bak}).

As far as the derivation of quantum theory as an \tit{effective
  theory} is concerned we therefore will assume that our network has
made a transition into this new phase, $QX/ST$, consisting, on a limit
scale of magnification, of a web of lumps, fluctuating around some
stabel positions and/or average shape with their degree of overlap and
their mutual (macroscopic) distance also fluctuating around some
average value. Some aspects of this picture are then very reminiscent
of model systems studied in the past (investigations initiated by
Menger et al; see e.g. \cite{Roy} and further
references given there). We sum up what we have said so far in the
following brief r\'esum\'ee. (Note that in the following, in contrast
to \cite{5}, we denote lumps or physical points by $P_i$ for
notational convenience).
\begin{ob}[The Two-Story Concept of QX/ST]\hfill\\
\begin{enumerate}
\item Given a network or graph, $G$, of the above kind, we can
  construct its associated {\em clique graph} $\mcal{C}_G$ (vertices
  being the lumps or cliques, the bonds given by overlap of cliques, see
  \cite{5}), 
and thus
  establish the two story concept, mentioned already in the
  introduction. We hence have two kinds of {\em distances} and {\em metric
    (causal) relations} in the network, the one defined by the
  original node distance in $G$, the other by the distance between
  lumps (defined by {\em overlap}) in $\mcal{C}_G$.
\item It is important that two lumps, $P_1,P_2$, which are some
  distance apart in $\mcal{C}_G$, may nevertheless be connected by a
  certain (possibly substantial) number of {\em interbonds} or short
  paths, extending from nodes in $P_1$ to nodes in $P_2$
 (see the construction of the cliques described in the preceding sections). 
\item That is, there may exist two types of information transport or
  correlation   being exchanged in the network. A relatively coherent
  (and presumably  {\em quasi-classical}) one, exchanged among the
  lumps, obeying a so-called {\em Nahwirkungsprinzip} and a more
  stochastic and less organized one (of {\em quantum nature})
  between individual groups of nodes lying in lumps, which may be a
  certain distance apart, and
  which, nevertheless, can be almost {\em instantaneous}.
\end{enumerate}
\end{ob}
  
As we are in the following mainly concerned with the information flow
between the various lumps, $P_i$, making up the orderparameter
manifold, $ST$, we develop below a couple of useful concepts and tools
which are adapted this new emergent level of description. Particularly
impotant for the \tit{near-/far-order} in $ST$ (which is related to
its \tit{causal structure}) are the various degrees of connectedness
among the physical points, $P_i$. The following abbreviation is useful.
For $n_i,n_k$ (not being) connected by a bond we write
\begin{equation}n_i\sim n_k\quad (n_i\not\sim n_k)\end{equation}
We then have
\begin{ob}From the definition of the cliques it follows 
\begin{enumerate}
\item $n_i\not\sim n_k$ implies that they are
lying in different $P_{\nu}$'s.
\item $P_{\nu}\,,\,P_{\mu}$ are disjoint, i.e. $P_{\nu}\cap
P_{\mu}=\emptyset$ iff 
\begin{equation} \forall n_{\nu}\in P_{\nu}\;\exists\; n_{\mu}\in
P_{\mu}\;\text{with}\;n_{\nu} \not\sim n_{\mu}\;\text{or vice versa}\end{equation}

\end{enumerate}
\end{ob}
This shows that it may well be that $P_{\nu}\cap P_{\mu}=\emptyset$
while the two lumps have still a lot of \tit{interbonds}, i.e. bonds
connecting the one with the other. The guiding idea is however that
the respective vertesx sets $V_{P_{\nu}}$ and $V_{P_{\mu}}$, as a
whole, will typically be weaker entangled with each other than the
nodes within $P_{\nu}$ or $P_{\mu}$ when the unfolding process is
fully developed.
\begin{obdef} With respect to the above clique graph or web of lumps we can
speak of an\\
\begin{enumerate}
\item {\em interior bond} of a given $P_{\nu}$, i.e:
\begin{equation} b_{ik}\;\text{with}\,n_i,\,n_k\,\in P_{\nu}\end{equation}
\item {\em exterior bond} with respect to a given $P_{\nu}$, i.e:
\begin{equation} b_{ik}\;\text{with}\,n_i,\,n_k\,\notin P_{\nu}\end{equation}
\item an {\em interbond}, i.e:
\begin{equation} b_{ik}\;\text{with}\,n_i\in P_{\nu},\,n_k\in
P_{\mu},\,\nu\neq\mu\end{equation}
\item a {\em common bond} of $P_{\nu}$,$P_{\mu}$ if $b_{ik}$ is an
interior bond both of $P_{\nu}$ and $P_{\mu}$.
\item a {\it true interbond} $b_{ik}$ if for $\nu\neq\mu$:
\begin{equation} n_i\in P_{\nu},\,n_k\in P_{\mu},\,n_k\notin P_{\nu}\end{equation}
\item We then have the relation for given $P_{\nu},\,P_{\mu}$:
\begin{equation}
  \{interbonds\}-\{common\;bonds\}=\{true\;interbonds\}\end{equation}
\end{enumerate}
\end{obdef}
We noted above that we now have two (metric) structures on the network
or graph, the original one with its neighborhood structure and
distance function, $d(n_i,n_j)$, and the superstructure given by the
clique graph and its coarse grained neighborhood structure of physical
points and coarse grained distance function, $d_{cl}(S_i,S_j)$, which
we regard as a protoform of our ordinary macroscopic distance. Note
that there may still exist a substantial number of interbonds on the lower
level between supernodes $P_i,P_j$ with $d_{cl}(S_i,S_j)\gg 1$.
  
In the physics of many degrees of freedom what really matters, or
gives ``distance'' a physical content, is not so much some abstract
notion of distance but the strength of interaction or correlation
between the various constituents. Given two node sets $A,\,B$ or the
respective subgraphs we can count the number of bonds connecting them
and regard this as a measure of their direct mutual dynamical
coupling.
\begin{defi}[Connectivity of Subgraphs] With $A,\,B$ being two node
  sets in a given graph, we denote by $|A\sim B|$ the actual number of bonds
  connecting the nodes of $A$ with the nodes of $B$ and by $|A\sim
  B|_m$ the maximal possible number. Then we call
\begin{equation} 0\leq c_{AB}:=|A\sim B|/|A\sim B|_m\leq 1\end{equation}
the {\em connectivity} of the pair $A,\,B$. It represents the
probability that a randomly chosen pair of nodes $n_A\in A,\,n_B\in B$
is connected by a bond.  $|A\sim B|_m$ depends however on their
relative position in $G$.
\end{defi}
\begin{ob}We have the following relations            
\begin{equation} A\cap
B=\emptyset \to |A\sim B|_m=|A|\cdot|B|\end{equation}
 ($|A|,\,|B|$ the
respective number of nodes), hence 
\begin{equation} c_{AB}=|A\sim
B|/|A|\cdot|B|\end{equation}
\begin{equation} A=B\to |A\sim B|_m=\binom{|A|}{2}\end{equation}
For their intersection we have in general 
\begin{multline} 
A\cap B\neq\emptyset\to|A\sim B|_m  =  |(A-B)\sim(B-A)|_m\\
                                    +|(A\Delta B)\sim(A\cap
                                   B)|_m+|(A\cap B)\sim(A\cap
                                   B)|_m
\end{multline}
i.e:
\begin{equation} |A\sim B|_m=|A-B|\cdot|B-A|+|A\cap
  B\cdot(|A-B|+|B-A|)+
 \binom {|A\cap B|}{2}\end{equation}
with $A\Delta B$ being the symmetric difference of $A$ and
$B$.
\end{ob}

In our papers, cited above, we argued that it is reasonable to treat
certain (bulk) aspects of the network properties and its evolution in
a statistical way. This is particularly necessary if we want to
extract some coarse-grained information from it which depends on the
collective behavior of many nodes and bonds. That is, we have to take
averages over certain portions of the network and/or (possibly
appreciable) clock-time intervals (which may, nevertheless, correspond
to infinitesimal intervals on a more macroscopic scale). In the
folowing we introduce and describe only those collective variables
which may become relevant in the further analysis.

We will deal with our network mainly on the level of the web of
lumps or the \tit{clique graph}, abbreviated by $ST$. At each
clock-time step or whole clock-time interval, $QX/ST$ consists of a
certain overlapping web of lumps or cliques, $P_i$, having some average size
(in graph theory usually called \tit{order}), that is, number of nodes
\begin{equation}\langle r(P)\rangle:=\langle
  order\;of\;clique\rangle\end{equation}
where the statistical average is taken over the network and/or an
appropriate clock-time interval. We assume of course that the phase,
$QX/ST$, the network or clique graph is occupying, is sufficiently
stable or slowly varying, so that the actual clique size, $\langle r(P_i)(t)\rangle$ is assumed to
fluctuate not too much around this average value, $\langle
r(P)\rangle$. In other words, the lumps are assumed to be \tit{fuzzy}. 

In the same sense the average vertex degree can be defined
\begin{equation}\langle v(P)\rangle:=\langle
  vertex\;degree\rangle\end{equation} 
the average number of active bonds per clique
\begin{equation}\langle N(P)\rangle:=\langle
  number\;of\;active\;bonds\;per\;clique\rangle\end{equation}
and the respective averages over the bonds pointing \tit{inward} or
\tit{outward}, that is, connecting two nodes, the one lying inside,
the other outside the lump under discussion.
\begin{equation}\langle N_{in,out}(P)\rangle:=\langle number\;of\;in-\;,\;out-bonds\rangle\end{equation}
Note that the bond is counted as \tit{in} if the \tit{orientation} is
$J_{ik}=+1$ with $n_i$ lying outside the lump. 

We consider these latter variables (averaged or non-averaged) as being
particularly relevant as they tell us something about the \tit{charge
  fluctuations} inside the lumps and through the (fuzzy) boundary. At each
clock-time step we have an internal charge, $Q(t;P)$, of the
respective lump, $P$, and due to our dynamical network law it holds
\begin{multline}Q(t+\tau;P)-Q(t;P)=:\Delta Q(t;P)=\sum_{n_i\in P}s_i(t+\tau)-
 \sum_{n_i\in
   P}s_i(t)\\
=q(N_{in}(t;P)-N_{out}(t;P))=q(\sum_{in}J_{ik}(t;P)-\sum_{out}J_{ik}(t;P))\end{multline}
and the corresponding equations when taken with the respective
averages.
\section{An Alternative Look upon Quantum Theory}
\subsection{Isolating the Pure ``Quantum Phenomenon''}
After this series of preparatory steps we now enter the central part
of the paper. We have our model system $QX/ST$ and want to derive
quantum theory from it as an effective theory living near the
continuous ``surface'' of this structure. In doing this we have first
to clarify two points. What do we actually mean by quantum theory
(understood as a general conceptual framework) and, second, what are
the large-scale phenomena we expect to emerge or survive in this
low-energy limit (comparedd to the primordial Planck scale).

As to the first question, typical models of, say, quantum field theory
are usually inspired by their classical counterparts which are then
``quantized'' (following a certain, one may venture to say heuristic,
scheme). Furthermore, \tit{covariance, spectrum condition etc.} are
usually imposed. In a first step we think it is easier to concentrate
on, what we regard as the essential and model-independent  quantum
phenomena, leaving, for the moment, aside all the additional
complications. Take e.g. \tit{special relativity}. To derive it as a
macroscopic phenomenon from our underlying $QX/ST)$-network we need a
more detailed understanding of the emergence of \tit{macroscopic
  time}, which is a veritable problem of its own and will not be dealt
with here (this does not mean that it cannot be done; it means rather
it has do be done in a separate investigation due to natural limits of
space), for reviews see e.g. \cite{time} and \cite{Butter}.
\begin{conj}We conjecture that the model-independent content of
  ``the'' quantum phenomenon is mainly encoded in the {\em generically
    complex structure} together with the {\em superposition
    principle}, both making up the {\em complex Hilbert-space
    structure} of ordinary quantum theory and leading to the
  seemingly paradoxical {\em entanglement phenomena}.
\end{conj}
Remark: The soundness of this conjecture will be further illuminated
below and is perhaps underpinned by the following citations taken from   
\cite{9} p.216, 218 (see also \cite{Dirac}):
\begin{quote}
{\small (Schr\"odinger)\ldots the complex structure as carrier of the
unobservable phase information\ldots\\
(Dirac)\ldots the phase quantity was very well hidden in nature\ldots}
\end{quote}
These model-independent ingredients can, in a first step, best be
studied in the non-relativistic regime by investigating the (hidden)
structure of e.g. the Schr\"odinger equation. 

In this context it is sometimes argued that some of the peculiar
quantum phenomena displayed by the Schr\"odinger equation, e.g. its
instantaneous spreading, is an artefact of its lack of
\tit{relativistic covariance}. We think, this is beside the point to
some extent. We carefully analyzed this issue in \cite{Req1} and
\cite{Req2} and showed that exactly the same processes are at work in
both the relativistic and the non-relativistic regime, pointing to a
kind of underlying \tit{entanglement} or \tit{non-locality} of quantum
phenomena, being ubiquituous in the whole field. Put sloppily one may
say:
\begin{conj}Only {\em observables} behave {\em locally} or {\em causally}
  while {\em states} do, typically, not. This is also the case in the
  relativistic regime. On the other side, the Schr\"odinger equation
  describes the evolution of a state!, {\em not} of a quantum field.
\end{conj}
The underlying reason for this is the following. The
\tit{energy-momentum content} or \tit{transfer} of an \tit{observable}
is typically \tit{two-sided}, that is, the Fourier-support or spectrum
(in fact an operator-valued distribution as long as the operator is
not smeared with a testfunction) 
\begin{equation}\hat{A}(p):=(2\pi)^{-2}\int
  e^{ipx}A(x)d^4x\end{equation}
$p=(p_0,\mbf{p}),x=(x_0,\mbf{x})$ four-vectors, $px$ the
Minkowski-scalar-product
\begin{equation}A(x)=e^{iPx}\cdot A\cdot e^{-iPx}\quad
  P=(H,\mbf{P})\end{equation}
has the following property:
\begin{lemma}With $p$ belonging to the energy-momentum support of
  $A=A^*$, put sloppily $\hat{A}(p)\neq 0$ at $p$ in a {\em distributional}
  sense, $-p$ belongs also to the support of $A$ or $\hat{A}$.
\end{lemma}
Proof: 
\begin{equation}(\hat{A})(p)^*=(\int e^{ipx}A(x)d^4x)^*=\int
  e^{-ipx}A(x)d^4x=\hat{A}(-p)\end{equation}
that is, $\hat{A}(p)\neq 0$ at $p$ as operator-valued distribution
implies the same for its adjoint at $-p$\hfill$\Box$\\[0.5cm]
Remark: In the theory of operator algebras the above spectrum is
usually called the \tit{Arveson-spectrum} (see e.g. \cite{Pedersen} or
\cite{Kastler}).\\[0.5cm]
States, on the other side, are prepared by applying such local
observables (or field operators) to the ground state or
\tit{vacuum}, $\Om$. As this is the state with lowest energy (usually set to
zero), the negative energy support of $A$, when applied to $\Om$, is
by definition cut off, that is,
\begin{equation}supp(A\cdot\Om)\subset V^+\end{equation}
($V^+$ the closed \tit{forward cone}). This inevitable asymmetry in
the support of states compared to observables leads to their different
(causal) behavior as has been analysed in our above mentioned papers
and has nothing to do with Lorentz-covariance. It is rather a pure
\tit{quantum phenomenon}.
\subsection{A Different Look at Schr\"odingers Equation}
In standard quantum theory the Schr\"odinger equation is considered to
be only one of a couple of possible representations of quantum
dynamics, i.e. the \tit{configuration-space} version. We want to argue
in this subsection that one should perhaps change ones point of view
in at least two respects. For one, as we learned from general
relativity, space and time seem to have a very peculiar significance
of their own and do not seem to be a mere mode of representation of
physics among many other ones being posible (in contrast to the point
of view, suggested by e.g. ordinary quantum theory). For another,
Schr\"odingers equation is at first glance linear (which was severely
criticized by Einstein), but it is only linear with respect to its
\tit{complex structure} which makes a big difference as we will show.

In our (perhaps heretic) view, shared however by quite a few others,
it rather represents an intricate \tit{dynamical entanglement} of two
underlying and in the ordinary approach largely hidden  quantities,
which have survived the coarse-graining process if one goes
``bottom-up'', starting from the Planck-scale. If one disentangles
this single complex-linear equation it becomes the non-linear coupled
evolution of two real equations. This fact has long been known, we
think however that our interpretation of this phenomenon is a
different one. 

The two perhaps most widely known fields where this has also been done
are the \tit{stochastic mechanics} developed by Nelson and several
precursors, and the so-called \tit{Bohmian mechanics} of Bohm et
al. To keep the length of our paper reasonable we will mention only
very few sources, where the interested reader can look up more
references, and make up his own mind concerning the pros and cons of
the various contributions. We mentioned already \cite{18} and
\cite{24} to \cite{26} which deal primarily with stochastic
mechanics. From the many papers about Bohmian mechanics we cite the
following, as we think, quite readable accounts \cite{39} to \cite{41}
with reference \cite{40} being perhaps particularly interesting as in
it Nelson's approach has been compared with Bohm's own approach.

In more recent times so-called \tit{stochastic collapse models} have
also become fashionable (a small selcetion being \cite{42} to
\cite{44}). As far as the randomly fluctuating environment, employed
in some of these models, can be viewed as a coarse-grained
epiphenomenon deriving from a more fundamental layer of microphysics,
they may considered to be phenomenological or effective theories,
describing a more complex underlying dynamics. Note however that in
our approach the fluctuating environment is a dynamical agens of its
own which acts but is also acted upon by ``matter''.Furthermore we do
not employ a \tit{particle picture} in the form of, say, small objects
immersed in a random medium. It may however has its value as a certain
approximation.

In disentangling the Schr\"odinger equation we follow a traditional
line of non-orthodox quantum theory but with a to some extent
diffferent working philosophy in mind, formulated in observation
\ref{geomatt}. That is, we conjecture that a really fundamental law
consists always of two parts, i) ``geometry'' acting on ``matter'' and
ii) ``matter'' acting on ``geometry'', with the first equation being
typically significantly simpler than the latter one. With
\begin{equation}\psi=\rho^{1/2}\cdot e^{iS/\hbar}\end{equation}
Schr\"odinger's equation
\begin{equation}i\hbar\partial_t\psi=-\hbar^2/2m\cdot\Delta\psi+V\psi\end{equation}
decomposes into the conservation equation
\begin{equation}\partial_t\rho=-\nabla\cdot(\rho\cdot\mbf{v})\;\text{with}\;\mbf{v}=1/m\cdot\nabla
  S\end{equation}
and the dynamical equation
\begin{equation}\label{Jacobi}-\partial_t S=1/2m\cdot(\nabla
  S)^2+V-\hbar^2/2m\cdot\Delta\sqrt{\rho}/\sqrt{\rho}\end{equation}
which may be considered as a \tit{quantum deformation} of the
\tit{Hamilton-Jacobi equation}, the deformation being the so-called
\tit{quantum potential} (Bohm)
\begin{equation}V_q:=-\hbar^2/2m\cdot\Delta\sqrt{\rho}/\sqrt{\rho}\end{equation}\begin{ob}The
  {\em quantum potential}, $V_q$, is the only place where the
  ``quantum'' openly enters. Thus, any attempt to explain quantum
  mechanics as arising from a more primordial level has to give an
  explanation for the emergence of this term.
\end{ob}
We will show that, in fact, this term encodes in our
framework the non-local entanglement between the various lumps making
up $ST$ in our network $QX/ST$.
\begin{bem}Note that in the conservation equation the quantity $\rho$
  enters, while in the second equation it is $\sqrt{\rho}$, the
  peculiar statistical nature of which will play a considerable role
  in the following sections.
\end{bem} 
There exist corresponding equations for several particles. For two
particles we have e.g.
\begin{equation}\label{one}\partial_t\rho=-(\nabla_1\cdot(\rho\nabla_1
  S/m_1)+\nabla_2\cdot(\rho\nabla_2 S/m_2))\end{equation}
\begin{multline}\label{two}-\partial_t S=(2m_1)^{-1}(\nabla_1 S)^2+(2m_2)^{-1}(\nabla_2 S)^2+V(x_1,x_2)\\-(\hbar^2/2m_1)\Delta_1(\sqrt{\rho})/\sqrt{\rho}-(\hbar^2/2m_2)\Delta_2(\sqrt{\rho})/\sqrt{\rho}\end{multline}

In the following conjecture we indicate how this disentangled
Schr\"odinger equation fits in our general picture.
\begin{conj}The conservation equation encodes the action of
  ``geometry'' on ``matter'', the deformed Hamilton-Jacobi equation
  the action of ``matter'' on ``geometry'', where $\rho$ is supposed
  to relate to ``matter'', the phase-function $S$ to
  ``geometry''. These structural elements survive the huge gap between
  the Planck-scale and the middle-energy regime of, say, quantum
  theory in the form of large scale {\em excitation patterns}.
\end{conj}

The role of the above quantities in ordinary quantum theory is the
following. We have
\begin{equation}\int \overline{\psi}i\partial_t\psi d^3x=-\int
  \rho\partial_t Sd^3x+i\int
  \rho^{1/2}\partial_t\rho^{1/2}d^3x\end{equation}
The latter term on the rhs equals $(i/2)\partial_t\int
\rho d^3x=0$. Hence $-\rho\cdot\partial_t S$ may be interpreted as an
\tit{energy density}. Correspondingly we have for the momentum:
\begin{equation}\int \overline{\psi}i^{-1}\partial_j\psi d^3x=\int
  \rho\partial_j Sd^3x+(2i)^{-1}\int \partial_j \rho d^3x\end{equation}
The last term on the rhs is a surface term and hence vanishes; thus
$\rho\cdot\partial_x S$ can be regarded as a \tit{momentum density}.

Before we go on we want to address again the longstanding question of
the reality of, say, the wave function $\psi$ or of its constituents
$\rho$ and $S$. This issue was already discussed in subsection 2.2 and
we now want to add a few more facets. One of the most prominent
scientists in favor of an undulatory ontological nature of $\psi$ was
Schr\"odinger (see e.g. his contribution in the de Broglie volume,
which contains quite a few remarkable observations supporting his
point of view). One argument against this interpretation is the
so-called problem of \tit{polydimensions}, i.e. the structure of
$\psi$ in the case of several particles, which is in that context
defined over the cartesian product of, say, $\R^3$ and hence can, at
first glance, no longer be interpreted as an extended excitation
living in $\R^3$. This point was already raised by Heisenberg at the
1927-Solvay conference (seee \cite{9} p.240f).
\begin{quote}
{\small
\ldots I see nothing in the calculations of Mr. Schr\"odinger that
justifies his hope that it will be possible to explain or understand
in three dimensions the results from polydimensions}
\end{quote}
While Schr\"odinger did not seem to have a coherent underlying theory
supporting his own point of view, he evidently envisioned excitations
in $\R^3$ interpenetrating each other (perhaps in a soliton-like manner).  
We will sketch our own ideas concerning this important question in a
preliminary form in the following conjecture.
\begin{conj}\label{several}[N-Particle Wave Functions]\hfill\begin{enumerate}
\item We think the {\em system-theoretic} task, solved by ``Nature''
  with the ``invention'' of quantum theory, consists of storing
  effectively and stably a certain amount of {\em information} in a
  {\em noisy background}, represented by $QX/ST$.
\item As the individual grains, $P_i$, i.e. the physical points,
  comprise in our picture still a lot of internal degrees or freedom
  (the nodes and bonds belonging to $P_i$), there should be ample internal
  space to store the local pieces of different excitation patterns,
  living and interpenetrating each other in one and the same
  emvironment (by the way, a task also solved by the human brain).
\item One possible method consists of letting only a small fraction of
  internal degrees of freedom contribute to each extended wave
  pattern. This appears to be reasonable anyhow, as quantum theory as we
  understand it is actually a weak, low-energy excitation of $QX/ST$ as
  compared to, say, the Planck energy.
\item An interesting situation is expected to emerge when the number
  of particles, $N$, becomes appreciable or macroscopic. There should
  exist a {\em critical local occupation rate} above which this
  weak-field-approximation breaks down. By the same token, the picture
  of interpenetrating (and to a certain extent individual) particle
  excitations will become   problematical. This impossibility to store
  a too large amount of   information in a finite space and the
  respective {\em threshold} are in our view the interface region
  where {\em quantum mechanical many-body systems} start to behave
  {\em macroscopically}. This picture has to be worked out in much
  more detail and will be treated elsewhere as it draws on a huge
  corpus of material of its own which has accumulated in the past (for
  a cursory discussion see however section 7).
\end{enumerate}
\end{conj}    
\section{The Collective Dynamics of the Web of Lumps}
In the preceding sections we described how a certain extended
structure of lumps may emerge within the network as the consequence of
a geometric phase transition or geometric reorganisation of the
underlying network. In this section we want to argue that this new
phase, dubbed $QX/ST$, is accompanied by the emergence of a new class of
characteristic \tit{collective variables} and their respective
\tit{cooperative behavior} which does not yet exist on the more
primordial level and which is the reason why this new geometric phase
may be rightly called an \tit{orderparameter manifold}.\\[0.3cm]
Remark: It is not accidental that such a point of view and/or language
is a little bit in the spirit of the physics of
\tit{self-organisation} or \tit{synergetics} (see e.g. \cite{Haken} or
\cite{Haken2}) as it is in our view pretty much to the point. After
all, our underlying medium is a complex \tit{dynamical system}
consisting of a huge number of elementary constituents. It is hence
reasonable to employ the corresponding arsenal of tools and
concepts. We will, due to limits of space, however only introduce a limited
amount of the technical machinery below.
\vspace{0.3cm}

The above described superstructure, $ST$, overlying the primordial
network, $QX$, is the deepest of, presumably, a whole hierarchy of
increasingly coarse-grained and smooth levels, each of which typically
supporting and generating its own emergent set of natural variables
and laws. In a sense $ST$ functions as a \tit{shell} which
\tit{decouples} and \tit{shields} the upper stories of the hierarchy
from the most primordial one. The crucial question to answer is the
following.
\begin{prog}Find the modes in which the system operates on a given
  scale of resolution!
\end{prog}
This is the characteristic question, emerging also in the physics of
self-organisation, the Landau-picture of elementary excitations in,
say, quantum fluids, or the theory of renormalisation in high-energy
physics, to mention a few fields.

\begin{conj}As to the cooperative behavior of our web of lumps we
  assume the following: The orderparameter manifold,
  $ST$, overlying the primordial network, $QX$, {\em enslaves} (a notion taken also from synergetics, see
  \cite{Haken}) the more primordial degrees of freedom, that means in
  our context the node- and bond-variables and forces them into a
  specific cooperative undulatory behavior, put differently, the {\em
    geometric phase transition} manifest itself, among other things,
  by means of a new collective excitation mode. In brief, order
  parameters are collective variables which enslave subsystems.
  
  This new collective mode is a {\em spatio-temporal} undulation
  pattern of the $Q(P)$- and the $N_{in,out}(P)$-field (see the end of
  section 3.2) being entangled with it via the underlying dynamical
  law. The emergent {\em characteristic parameters} of this excitation
  mode are an {\em oscillation-} or {\em correlation time}, $t_{pl}$, and
  a characteristic {\em (correlation- or oscillation-) length}, $l_{pl}$,
  supposed to characterize the Planck-regime.
\end{conj}
While we cannot prove this conjecture at the moment, as we are
presently unable to solve the very complicated dynamical behavior of
our network in greater detail and follow it through its presumed phase
transition regime into the new phase, $QX/ST$, we will at least try to transform it into an \tit{educated
  guess} by providing a row of more or less coherent arguments
supporting this hypothesis.
\begin{enumerate}
\item We mentioned in section 3 in the remark after the definition of a, as we
  think, typical model of a dynamical network law (definition
  \ref{law}) that a slight variation of the law yields a new model
  having very short \tit{transients} and reaches \tit{periodic
    state cycles} or (\tit{attractors}) having only period six, i.e. the whole network state
  returns into exactly the same state after only four time steps. For
  sufficiently small networks (only a few nodes) one can do the
  calculations by hand and follow the evolution and oscillation in
  detail. For networks up to several thousand nodes the evolution has
  been implemented on a computer (see \cite{Nowotny}). Note that such a behavior 
  is quite remarkable, given the huge acccessible
phase space and the relatively complicated evolution law.\\[0.3cm]
Remark: Such puzzling and still quite mysterious phenomena were also
mentioned by S.Kauffman in his study of so-called \tit{switching nets}
(cf. e.g. \cite{Waldrop} p.112 ff or \cite{Kauffman}). Such
\tit{oscillating media} are also observed in synergetics
(cf. e.g. section 4 in \cite{Mikhailov1}). 
\\[0.3cm]
This shows that such things may already happen on the level of
the primordial network. In other respects however,such a peculiar law is too
regular as it does not allow for a diversified pattern creation on the higher
levels. The interesting evolution laws are sitting, according to the
working philosophy expounded e.g. in the above cited literature (see
also \cite{Bak}) and which we
are largely sharing, at the edge between \tit{chaos} and \tit{order}.
\item On the other side, the evolution law given in definition (\ref{law})
  itself, i.e. with the other rule of switching-on and -off of
  interaction, $J_{ik}$, does not show these short transients
  and periodic state cycles already on the level of its primordial
  nodes and bonds    and therefore
  seems to behave more erratically at least on the most fundamental
  level (as we learned from a numerical investigation of
  its characteristics which were also studied in quite some detail in
  \cite{Nowotny}). It evidently behaves more stochastically and may be
  nearer to this mentioned edge between order and chaos (sitting
  supposedly on the other side). Unfortunately
  our computer capacities were not large enough to study it on the
  higher level of the web of lumps., which would have implied, among
  other things, a permanent calculation of these cliques and their
  dynamics (as to such analytical techniques cf. \cite{5}). It may,
  however, be possible that it supports such oszillating modes in its
  fluctuation spectrum on this more advanced level of lumps.
\item On can however try to get some qualitative glimpses how such
  networks may behave by concentrating on a fixed given lump, $P$,
  say. Let us assume that at a certain clock-time, $t_0$, its charge,
  $Q(t_0;P)$, happens to be somewhat below the average of the charges
  of the surrounding lumps directly interacting with it (i.e. more or
  less the so-called \tit{local group}, see \cite{5}). While we cannot make
  an exact prediction about the corresponding states of the bonds in
  the immediate environment of $P$, the laws which we introduced above
  suggest however that a deficiency of charge in $P$ will induce after
  some clock-time cycles a reorientation of bonds in favor of more
  bonds pointing inward (with, and this is important, a certain tendency of
  \tit{overcompensating}!). In other words, after a certain lapse of
  time the charge in $P$ will be above average with the excess charge
  now being presumably greater than the deficiency of charge in the
  cycle before. Again this surplus charge induces an even more
  pronounced reorientation of neighboring bonds, leading to an even
  greater deficiency of charge within $P$, and so on. This
  process will not stop until it has reached a
  \tit{characteristic excitation level} being typical for both the network
  (law) under discussion and the specific phase, $QX/ST$ it is occupying.
\end{enumerate}

The above qualitative analysis shows that there may well be such a 
spatio-temporal {\em undulation pattern} within the excitation spectrum of the
  network, extending over the whole web of lumps and being perhaps
  similar to an array of coupled self-oscillating subunits. The
  characteristic {\em oscillation period} of these subunits (which are
  assumed to comprise more or less the individual lumps and there
  immediate neighborhoods) is the {\em Planck-time}, $t_P$, the
  characteristic {\em wave-} or {\em coherence length} is the {\em
    Planck-length}, $l_P$, being, on the other side, a measure of the typical
  diameter of a lump or its local neighborhood. These emergent and
  autonomously generated quantities figure then as the elementary
  building blocks of the corresponding continuum concepts, length,
  time, energy etc. on the smoother, that is, more coarse-grained  scales and show how a complex system
  is capable of generating its own intrinsic scales by a process of
  {\em self-organisation}. That is, these elementary units need not be
  put in by hand! That is, our subclass of networks seem to belong to
  the class of \tit{oscillating media} described in e.g. \cite{Mikhailov1}.
  
  It seems now worthwhile to introduce a limited amount of machinery,
  being employed in the theory of self-organization or dynamical
  systems (see e.g. \cite{Haken} to \cite{Ott}), in order to make the
  following analysis more concise. Our networks are, among other
  things, complex dynamical systems. On the most fundamental level
  their dynamics is assumed to be \tit{deterministic} (whereas this is
  not neccessarily a crucial point due to the \tit{shielding
    phenomenon} which decouples the various levels from each other to
  some extent). In contrast to most of the dynamical systems,
  discussed in the literature, the number of their constituents is, on
  the one side, very large. On the other side, both the evolution and
  the phase space is \tit{discrete}. This prevents the immediate
  application of the usual tools of continuum mathematics in the
  analysis of the geometry of, say, \tit{attracting sets} and the
  like. On the other side, after a certain coarse-graining, when
  dealing e.g. with the web of lumps, the dynamics and the medium may
  be considered, in a good approximation, to be quasi-continuous. On
  the other side, on the higher levels the dynamics is no longer
  deterministic due to the ``integrating out'' of degrees of freedom
  and corrsponding loss of information. Instead of that we may get
  certain phenomenological dynamical field equations superposed by a
  stochastic component implementing the additional noise in the network. This then is the typical scenario of
  synergetics. It is our aim to show that low-energy quantum theory is
  exactly of this sort, that is, certain emergent dynamical field laws
  plus a \tit{non-local} stochastic component.

To exhibit the interplay of statistical averaging and the underlying
microscopic evolution, we introduce the folowing concepts. We take a
particular initial state, $x_0$, say, which lies in the \tit{basin of
  attraction} of an attracting set of states, $X$, say, or already in
$X$ itself. We assume a discrete evolution law, i.e. an \tit{iterated
  map}, $M$. In general, $M$ is not \tit{invertible} but only an
\tit{endomorphism}. For technical reasons it is usually assumed that
it is \tit{onto}. Then we can follow the path the system takes with
starting point $x_0$, i.e:
\begin{equation}x_n:=M^nx_0\quad , \quad M^n:=M\cdots M\;\text{($n$-times)}\end{equation} 
For $n\to\infty$ the states, $x_n$, wander through the attractor,
$X$. With $f$ an observable, defined on the microstate $x$, we can
define its \tit{time average} (provided it exists):
\begin{equation}\overline{f}:=\lim_{T\to\infty}1/T\sum_{n=0}^T
  f(M^nx_0)\end{equation}
Under certain conditions there exists an \tit{invariant measure}, $\mu$, on
$X$ so that time averages become ensemble averages with respect to
$\mu$, the averages being independent of the starting point,
$x_0$. Systems with this property are called \tit{ergodic}
(cf. \cite{Sinai}). How this may come about can be seen as
follows. Instead of $x_0$ we take a full initial probability
distribution, $\rho_0$. Under the map $M$ it goes over in a new distribution, $\rho_1$:
\begin{equation}M:\;\rho_0(x)\to \rho_1(x)\end{equation}
and in general
\begin{equation}\rho_{n+1}=\int \rho_n(y)\cdot
  \delta(x-My)dy\end{equation}
(called for whatever reason the Frobenius-Perron-equation).

We arrive at an \tit{invariant density} if we have a \tit{fixed
  point}, that is:
\begin{equation}\rho_{n+1}(x)=\rho_n(x)=\rho(x)\end{equation}
In general it is reasonable to switch to a slightly more general point
of view and consider \tit{invariant measures} instead of densities (or
to allow for distributional densities), that is $\rho(x)\to \mu(A)$
with $A$ some measurable set. The notion of invariance is now
expressed as
\begin{equation}\mu(A)=\mu(M^{-1}A)\quad\text{for all measurable sets}\end{equation} 
We then have
\begin{equation}\langle f\rangle=\overline{f}:=\lim 1/T\sum_{n=0}^T
  f(M^nx_0)=\int_X f(x)d\mu\quad\text{independent of
    $x_0$}\end{equation}

While $M$ is not necessarily invertible, invariance of $\mu$ implies
that it induces an \tit{isometric} map on the function space (Hilbert
space) $L^2(\mu)$, thai is, it holds
\begin{equation}\int |U_Mf|^2d\mu=\int|f|^2d\mu\quad\text{with}\quad
  (U_Mf)(x) :=f(Mx)\end{equation}
For this to make sense the above mentioned technical property that $M$
is onto is needed.
\begin{ob}Note that $\mu(A)$ measures in effect the average time the
  system occupies states belonging to $A$.
\end{ob}

Another useful tool in the analysis of such complex dynamical systems
is the method of the \tit{correlation functions} and their
\tit{spectral representation}. Suppose again that $f(x)$ is an
observable defined on our state space. Its \tit{time-autocorrelation
  function} is defined by
\begin{equation}\langle f(t_1)\cdot f(t_2)\rangle^C:= \langle
  f(t_1)\cdot f(t_2)\rangle -\langle f\rangle^2=\lim
  1/T\sum_{t=0}^T(f(t+t_1)-\overline{f})\cdot(f(t+t_2)-\overline{f})\end{equation}
(provided that such limits do exist). Fourier transformation leads to 
\begin{equation}\langle f(t_1)\cdot f(t_2)\rangle^C=(2\pi)^{-1/2}\cdot\int
  e^{-i(t_1-t_2)}\cdot c(\omega)d\omega\end{equation}
with $c(\omega)d\omega$ a positive measure, the so-called \tit{power
  spectrum} of the respective observable. If we have an invariant
measure these correlations can alternatively be calculated in the
ensemble approach (note that, as the evolution is usually
\tit{dissipative}, that is, one-sided or not-invertible, $t_1,t_2$ have to be chosen positive).

The power spectrum can be used to characterize the type of evolution
or/and attractor. If there are e.g. sharp peaks in the spectrum they
signal the existence of extended oscillating modes, buried in the
(possibly continuous) background noise. This is exactly the situation
we are speculating about in the case of our networks and the
particular phase, $QX/ST$, which is, in the language of dynamical
systems, an attractor. What we will say in the following about the
qualitative dynamics of our network or the web of lumps overlying it,
should be considered within this wider context which we only briefly
sketched above. One should however note that our networks are far more
complex than the dynamical systems usually considered in the
corresponding literature. For the time being we simply have to assume
that our phase $QX/ST$ corresponds to an attracting set, that
long-time averages are practically independent of the initial
configuration and that an invariant measure exists on the attractor,
corresponding to $QX/ST$, so that time averages can be expressed by
averages with respect to this measure. This lays the basis for a
\tit{statistical treatment} of the problems to be discussed in the
following.

Our qualitative discussion of the propensity for an oscillating
behavior of our medium (the web of lumps) suggests that we will find a
sharp peak (actually two as the spectrum is symmetric) in the Fourier spectrum of the (clock-)time autocorrelation
function of the  charge, $Q(P,t)$, of a given fixed lump, $P$. That is, we conjecture
\begin{equation}\langle Q(P,t)\cdot Q(P,t+t_1)\rangle^C=
  mode(\omega_{pl})+\int remainder\end{equation}
The spatio-temporal excitation pattern, resulting from the cooperation
of these individual resonating lumps, will presumably be much more
complex. It could be tested via the correlation among different lumps,
i.e:
\begin{equation}\langle Q(P',t')\cdot Q(P,t)\rangle\end{equation}
As we want to concentrate in this investigation on the derivation of
low-energy quantum theory, we will postpone a more detailed discussion
of the leading collective modes being prevalent in the vacuum on the
Planck-scale. We will only briefly indicate what kind of excitation
patterns we are having in mind, as it shows that our web of lumps may
already contain the so-called \tit{string-bits}, i.e the prerequisites
to allow for string-like excitations as cooperative patterns made of
local clusters of lumps (\cite{Thorn1}).

One possible excitation pattern may have the structure of densely
entangled \tit{chain mail}, built from elementary loops (the
chain-links) consisting on their side of lumps which resonate in a
cooperative manner so that a certain amount of charge is pulsating
around the respective loop. It is therefore perhaps not too
far-fetched to tentatively associate the lumps with the notorious
$D0-branes$ and their entangled interaction being modelled by a
\tit{matrix-model} (see
e.g. \cite{Susskind},\cite{Polchinski}).\\[0.3cm]
Remark: We denote in the following these presumed local clusters of
cooperating lumps by $C_i$.\vspace{0.3cm}

What is important for our further discussion is that the
\tit{wave-number}, $k_{pl}$, the pendant of $\omega_{pl}$, should not
be associated with some plane-wave excitations. The relevant
\tit{normal modes} are rather such entangled collective excitations as, say,
the above mentioned chain mail. The characteristic parameter,
$k_{pl}$, is then just the dual of $l_{pl}$, which, on its side,
characterizes the diameter of the local resonating lumps or the
elementary patterns being built from them, that is the, the local
clusters $C_i$.  

The characteristic parameters of our web of lumps are related to each
other as follows. We assume that the characteristic frequency is the
\tit{Planck-frequency}, the characteristic wave-number, $k_{pl}$, the
\tit{Planck-wave-number} etc. These variables are related with each
other via:
\begin{alignat}{2}
E_{pl} &  =\hbar\cdot\omega_{pl} & \qquad p_{pl} &  =\hbar\cdot k_{pl} \\
E_{pl}\cdot t_{pl} &  =\hbar & \qquad p_{pl}\cdot l_{pl} &  =\hbar \\
t_{pl} &  =l_{pl}/c & \qquad l_{pl} &
  =(G\hbar/c^3)^{1/2}
\end{alignat}
with $E_{pl},p_{pl},t_{pl},l_{pl}$ Planck-energy, -momentum, -time,
-length respectively. The remaining contributions in the spectrum we
assume to be a cetain amount of patternless noise plus longer
wave-length modulations of this \tit{ground oscillation} with typical
wave-length $l_{pl}$ and oscillation-time $t_{pl}$ (see the next
section).

To simplify the following discussion and to exhibit the red thread, we
neglect, in a first step, all the stochastic fluctuations and possible
modulations of this ground wave and concentrate on the leading mode
contribution. That is, we write for a fixed but arbitrary lump
(suppressing for the moment an additional phase factor)
\begin{equation}Q(P,t)\approx Q_{av}+Q_0\cdot \cos (\omega_{pl}\cdot
  t)\end{equation}
$Q_{av}$ is the average charge of the lump under discussion, which we
assume to be the same over the web of lumps, $Q_0$ is the amplitude of
the oscillation. In the following section we have to deal with our
network and/or the overlying web of lumps on several clearly separated
scales. The same holds for the respective natural observables emerging
on the various scales. The scale of our web of lumps we assume to be
associated with the Planck-scale, abbreviated by $[L_{pl}]$. We
further introduce the scale of ordinary quantum theory, denoted by
$[L_{qm}]$, with the property
\begin{equation}[L_{qm}]\gg[L_{pl}]\end{equation}
On the scale $[L_{pl}]$ we can incorporate the additional
spatio-temporal fine structure, induced by the resonating local
clusters, $C_i$, of the above described ground mode by introducing a
space-time dependent phase-factor, $\varphi(x,t)$, which varies over
the local clusters of lumps forming the elementary building blocks of,
say, the above mentioned chain mail. As it varies spatially on these very
short scales, that is, $l_{pl}$, it is almost invariant on average on larger scales as
e.g. $[L_{qm}]$ and can practically be ignored on this larger
scale. Furthermore, it should not change the imposed frequency,
$\omega_{pl}$ in an appreciable way. We hence expect:
\begin{equation}\partial_t\varphi\ll\omega_{pl}\quad ,\quad
 |\partial_x\varphi|=\mcal{O}(k_{pl})\end{equation}
\begin{assumption}[Fine Structure of the Physical Vacuum]
\begin{equation}Q(x,t)=Q_{av}+Q_0\cdot\cos(\omega_{pl}t+\varphi(x,t))+\;\text{noise}\end{equation}
\end{assumption}
\section{Quantum Theory as a Low-Energy Limit of the Dynamics of the
  Web of Lumps}
\subsection{The Building Blocks}
We now embark on the derivation of the building blocks of
\tit{low-energy quantum theory} as coarse-grained quantities from our
web of lumps. Most important are the two quantities $\rho$ and
$S$. From the \tit{quantum-Hamilton-Jacobi equation} of section 4 we
surmise that $S$ will play a particularly significant role as a
unifying concept, mediating between the Planck-, the quantum- and the
classical regime.

We begin with some heuristic considerations concerning the supposed
role of $S$ as mediator between these very different scales. Adopting
so-called \tit{natural} units with $c$ and $\hbar$ chosen
dimensionless, the phase function $S$ becomes a dimensionless quantity
which should be considered as a sort of \tit{generalized action}. On
the one side our working philosophy is that physics is discrete on the
fundamental level. On the other side we want to interprete quantities
like $S$, occurring in ordinary quantum theory, as something which
really exists. This leads to the following conjecture.
\begin{conj}The phase function $S$ describes the coarse-grained effect
  of an undulation phenomenon on the level of the web of lumps. More
  precisely, it counts the (dimensionless) number of oscillations or
  switches of or within the underlying fundamental medium with respect to a fixed but arbitrary reference point.
$-\partial_t S$ and $\nabla S$ have the dimensions of energy and momentum
in natural units, that is, inverse time and inverse length or
frequency and wave number characterizing the undulations in this
presumed substratum.
\end{conj}
\begin{ob} Note that $\rho$ is intrinsically positive in ordinary
  quantum theory while the phase function $S$ can be positive or
  negative. This makes no problems in the ordinary interpretation
  where $\rho$ is a probability density and $S$ has a relatively
  fictitious meaning. If we want to attribute some ontological meaning
  to them the situation is different (see below). It turns out that
  this ``{\em problem of signs}'' may be used as a guiding principle
  in isolating the relevant quantities which survive the
  coarse-graining process (cf. the related ``{\em problem of
    scales}'', discussed in section 2.3)
\end{ob}

The collective excitation pattern, we have described so far, will now
serve as the \tit{carrier wave} of smooth long-wave-length
undulations, which \tit{modulate} the ground excitation pattern and
which vary on their natural scale, $[L_{qm}]$ . In
other words, we will relate the objects of, say, \tit{low-energy
  quantum theory} to certain \tit{low-frequency/long-wave-length
  modulations} of this underlying very-high-frequency oscillation
mode. 
\begin{assumption}[Quantum Theory]
  Staying within our approximative picture, we expect the following
  modulation of the dominant mode in the carrier wave if there are
  low-energy quantum objects around. These quantum objects are assumed
  to be implemented by
  modulations of small \tit{amplitude} and \tit{frequency} (compared
  to the Planck-characteristics, $Q_0$ and $\omega_{pl}$). That is
\begin{equation}\label{quantum}Q(P,t)\approx Q_{av}+(Q_0+a(P,t))\cdot\cos(\omega_{pl}\cdot
  t+\varphi(P,t)+\varepsilon(P,t))\end{equation}
with 
\begin{equation}a(x,t)\;,\;\varepsilon(x,t)\;\text{varying on
    scale}\;[L_{qm}]\gg [L_{pl}]\end{equation}
and
\begin{equation}|a(P,t)|\ll
  Q_0\;,\;\partial_t\varepsilon(P,t)\ll\omega_{pl}\end{equation}
\end{assumption}   
In other words, a quantum object is assumed to consist of some
extended information pattern, being impressed on the high-frequency
carrier wave, representing on its side the \tit{physical vaccum}. This
impressed information consists of both a component implemented as
\tit{amplitude modulation} (i.e., $a(P,t)$) and a component being
realized as \tit{phase-modulation} (that is, $\varepsilon(P,t)$).

We interprete this modulation pattern in the following way on the
level of lumps. A space-time dependent tiny fraction, $a(P,t)$, of the
nodes within the given lump, $P$, joins the number of nodes (is slaved!), $Q_0$,
oscillating collectively (in the case $a>0$) or leaves this set, i.e.
falls out of phase (for $a<0$). At the same time the \tit{momentary
  frequency} is also changed by a tiny amount,
$\partial_t\varepsilon$.
\begin{obdef}We now relate $|a(x,t)|$ with $\rho(x,t)$ and
  $\varepsilon(x,t)$ with $-S(x,t)$. $(\omega_{pl}\cdot t)$ counts, as
  kind of a {\em generalized action}, the (dimensionless) number of
  oscillations or switches with respect to some arbitrary reference
  time (or -point). $\varepsilon(x,t)$ or $-S(x,t)$ is the local
  deviation from this global quantity (measured on scale $[L_{qm}]$),
  induced by the presence of quantum objects.  It is satisfying that
  $a(x,t)$ can be both positive or negative in contrast to
  $\rho(x,t)$. We assume however that it is either (in this low-energy
  approximation) positive or! negative, as we should expect both {\em
    particle-} and {\em hole-excitations} in our medium. These may
  correspond (in the old but perhaps not outdated Dirac-picture) to
  particles and anti-particles. The detection of the presence of an
  excitation (or particle) in low-energy quantum theory should, on the
  other side, not depend on the sign of $a(x,t)$.
\end{obdef}

With the help of the above formula (\ref{quantum}) we can now give
$-\partial_t S$ and $\nabla S$ a precise microscopic meaning.
\begin{ob}$-\partial_t S$ describes the local deviation of the
  momentary frequency of the undulation pattern from the vacuum value,
  $\omega_{pl}$, on scale $[L_{qm}]$. $-\nabla S\cdot dx$ measures on
  the one side the differential change of the number of periods of
  oscillation with respect to the reference point, $\omega_{pl}\cdot
  t$. On the other side, in order to be a physical observable, it must
  also have a meaning which can be locally measured in the
  medium. $\varphi(x,t)$ varies spatially on scale $l_{pl}$, i.e. it
  yields a wave number of order $k_{pl}$. $-\nabla S\cdot(dx/|dx|)$ is
  the local deviation of the wave number from $k_{pl}$ in direction
  $dx$, measured on scale $[L_{qm}]$. $\nabla S$ itself points in the
  direction of the maximal decrease.
\end{ob}
Proof of the latter statement: Take a change of phase of $2\pi$ on
scale $L_{qm}$ in direction $dx$. We have
\begin{equation}2\pi=\nabla S\cdot dx=\nabla
  S\cdot(dx/|dx|)|dx|\end{equation}
Hence
\begin{equation}\lambda=|dx|=2\pi(\nabla S\cdot
  dx/|dx|)^{-1}\;\Rightarrow\;k=2\pi/\lambda=\nabla S\cdot
  dx/|dx|\end{equation}
\subsection{The Quantum Mechanical Continuity Equation}
We now come to the derivation of the two defining equations making up
quantum theory in e.g. the Schroedinger picture. For one particle we
have
\begin{equation}\partial_t\rho=-\nabla(\rho\cdot\nabla
  S/m)\end{equation}
For a particle excitation we assume $\rho(x,t)=a(x,t)\ge 0$. Gauss-law
yields
\begin{equation}\partial_t\int_V a(x,t)d^3x=-\int_{\partial
    V}(a(x,t)\cdot\nabla S/m)d\mbf{o}\end{equation}
The amount of surplus charge, $\int a(x,t)d^3x$, activated by the
presence of the quantum excitation in the vacuum and participating in
the collective motion, is conserved in time. $\rho\cdot\nabla S$
 is a momentum density, $\rho\cdot\nabla S/m$ a ``velocity
 density''. We want to understand the above first law
 microscopically. 
 
 The above equation, interpreted on the level of the web of lumps,
 tells us that the surplus charge in $V$ has the tendency to move in
 the direction of decreasing wave number, $\nabla S$, or longer
 wavelength, with a prefactor, $m^{-1}$, which may be considered as a
 measure of the stiffness of the excitation against change. This seems
 to be a reasonable behavior.
\begin{bem}$m$ was already detected in classical mechanics as a measure
  of resistance of particle motion in the background medium called vacuum. The above
  microscopic interpretation seems to be consistent with this
  observation.
\end{bem}
Increase of wave number means shorter wave length, i.e. smaller
extension of the local clusters, $C_i$, in the direction of $\nabla
S$. This may imply a more intense coupling among the lumps cooperating
in the respective $C_i$'s, which may have the effect that more
elementary charges, $q$, are participating in the cooperative
movement, hence, an increase of $\int_V a(x,t)d^3x$ in the volume
$V$. That is, the first equation of low-energy quantum theory may be a
reasonable formula also on the microscopic scale.

We want to add at this place a comment about the \tit{normalisation
  condition} of ordinary quantum mechanics as a probabilistic
theory. In our realistic approach, in which $\rho$ is not considered
as some probability density but as kind of an (abstract) amount of
charge or information per lump, participating in a collective
undulating motion, a conservation law like $\int \rho d^3x=const$ may
be reasonable. On the other hand, a normalisation to, say, $const=1$
does not make physical sense in this more general non-probabilistic
framework. What may, however, be reasonable is a (\tit{projective})
\tit{ray-interpretation} as it is sometimes employed anyhow in quantum
theory (as to this more geometric aspect cf. the interesting paper by
Ashtekar and Schilling; \cite{Ashtekar}). 

In the same sense as mentioned before (cf. section 3.1), physics on
this primordial scale may be independent of the absolute values of
node-charges in this \tit{weak-field regime}. What may rather matter
is the relational information content being stored in the shape of the
excitation pattern. This idea does however not contradict the strong
probabilistic flavor of ordinary quantum theory. In this latter
framework probabilities are closely linked with observations and
measurements. There outcomes, on the other side, are of course related
to the information content of the excitation patterns under discussion
but this relation may be a subtle one (see section 7).
\subsection{The Quantum-Hamilton-Jacobi Equation, the Local
  Contribution}
More demanding is the interpretation of the second equation of
low-energy quantum theory. This applies in particular to a microscopic
understanding of the so-called \tit{quantum potential}, which contains
the core of quantum behavior. We begin with the first (drift-)term on
the rhs of equation (\ref{Jacobi}). Our working philosophy is that the
quantum potential contains the non-local stochastic effects while the
other terms encode the local and more coherent contributions.

Our analysis in the preceding sections shows that a local change (on
scale $[L_{qm}]$) of the \tit{action}, $S(x,t)$ relative to 
\begin{equation}S_0(x,t):=\omega_{pl}t\;\text{or}\;(\omega_{pl}t+\varphi(x,t))
\end{equation}
is accompanied by a change of local wave-number, $k(x,t)$. As we have
a coherent web of elementary oscillating circuits, a variation of
$k(x,t)$, that is, $\nabla S(x,t)$, will, by the same token, induce a
local change of frequency, $-\partial_t S(x,t)$. Hence there must
exist a \tit{dispersion law}
\begin{equation}\partial_t S=F(\nabla S)\end{equation}
(we neglect, for the moment, the other contributions in
(\ref{Jacobi})).

In the non-relativistic regime we can relate $d\omega/dk$ to a
velocity. The role of velocity in our context is played by $\nabla
S/m$. Hence, relating $-partial_t S$ with $\omega$ and assuming a
power-law behavior, we can infer
\begin{ob}
\begin{equation}-\partial_t S=(2m)^{-1}(\nabla S)^2\end{equation}
\end{ob}
On the other side, the potential term, $V(x)$, encodes some external,
effective force and is model dependent. The really crucial
contribution is however the quantum potential, $V_q$.     
\subsection{The Quantum Potential}
We now come to the most mysterious term in the (re)interpretation of
the equations of low-energy quantum mechanics, that is, the quantum
potential
\begin{equation}V_q=-(\hbar^2/2m)\cdot\Delta(\rho^{1/2})\cdot
  \rho^{-1/2}\end{equation}
We will argue that $V_q$, as non-relativistic quantum theory is still
relatively near to the classical regime, being perhaps only a small
deformation (compared to the Planck scale), is the only component
comprising truely stochastic elements in the above representation. By
the same token, it is the term being responsible for the seemingly non-local
phenomena, being almost ubiquituous in quantum theory. We will phrase
it that way:
\begin{conj}The quantum potential, $V_q$, encodes the {\em non-local}
  aspects of quantum theory in a, superficially, {\em local} way.
\end{conj}
\begin{bem}This non-locality is more hidden in the one-particle
  situation or the {\em self-interaction} among the terms within the excitation
  pattern belonging to a specific quantum object. It becomes more
  apparent when several quantum objects are involved. On the other
  hand, the interpretation of the meaning of the physical quantities
  becomes much more subtle in the latter case ({\em polydimensions});
  cf. secction 7.
\end{bem}

If we ignore this term, we have sort of a classical field theory (as
is of course well-known). In the free case ($V=0$) a solution of
\begin{equation}\partial_t\rho=-\nabla\cdot(\rho\nabla
  S/m)\;,\;-\partial_t S=(2m)^{-1}(\nabla S)^2\end{equation}
is
\begin{equation}\rho(x,t)=f(x-vt)\;,\;S(x,t)=mvx-mv^2t/2\end{equation}
with $\nabla S=v$, that is, $\rho$ spreads ``causally''; with
$\rho(x,0)$ of compact support it remains so for all $t$, being merely
shifted with velocity $v$. This is in sharp contrast to the quantum
case, i.e. after addition of $V_q$. We remarked in section 4.1 that we
typically observe an instantaneous spreading of (quantum) information
irrespectively of the details of the model (for the non-relativistic
regime see in particular \cite{Req2}). We emphasize again that this
ha, in our view, nothing to do with the relativistic non-covariance of
ordinary quantum theory but represents rather a (the) pure ``quantum
phenomenon''. 

When scrutinizing the structure of $V_q$ and having a stochastic
interpretation in the back of ones mind, two intriguing features
strike the eye.
\begin{ob}First, the prefactor, $\hbar^2/2m$, has the dimension of a
  {\em diffusion coefficient}, i.e. $[l]^2/[t]$, when energy is
  identified with frequency.

Second, if $\rho$ is some statistical sum (or average) over a
relatively large number of more primordial degrees of freedom,
$\rho^{1/2}$ may just describe the {\em standard deviation} or {\em
  typical fluctuation} of the additive quantity, $\rho$, about its
average.
\end{ob}
Then, the occurrence of the quantum potential in the
quantum-Hamilton-Jacobi equation tells us that the local momentary
frequency of the undulation pattern is influenced by a diffusive
and/or fluctuation contribution. We undertake to clarify the nature of
this term in two steps. First, we try to explain its very
existence. In a second step we try to provide an argument why it
changes the local frequency.

We argued in the preceding sections that, in our view, the apparent
non-locality of quantum theory has its origin in the two-storey
structure of the medium $QX/ST$, that is, the physical vacuum and
expresses itself in the (as yet not very well understood) complex
superposition principle of ordinary quantum theory. The terms we have
discussed so far were of a local nature and can be understood already
on the level of the web of overlapping lumps, forming the ``surface
structure'', $ST$, of our network $QX/ST$.

We noted however, that the lumps (that is, on a level of lesser
resolution, the physical points), even if they are non-overlapping (in
other words, being some distance apart in macroscopic space), may
nevertheless be connected in the underlying network by some (perhaps
even appreciable) number of \tit{interbonds} (cf. section 3.2). Via
these interbonds information may be exchanged almost
instantaneously. We expect however that this type of information
exchange is less \tit{coherent} and less \tit{organized}, that is,
more \tit{stochastic} than the information exchange taking place among
directly overlapping lumps, i.e between \tit{infinitesimally
  neighboring} regions of macroscopic space. Our central conjecture is
now that the mysterious quantum potential is just the remnant on scale
$[L_{qm}]$ of this kind of non-local coupling between non-neighboring
lumps or points!

We may consider each given lump or the respective local cluster, $C$, as
an \tit{open statistical subsystem}, being embedded in the ambient
space, $QX$, acting as kind of a \tit{reservoir}. Each of these lumps
or local clusters oscillates in its respective local coherent mode,
described above, having an amplitude, $Q_0+\rho(x,t)$ with the
deviation, $\rho(x,t)$ varying on the larger scale $[L_{qm}]$. Being a
sum over a relatively large number of more elementary degrees of
fredom we expect this deviation of the ground oscillation, $\rho$, to
fluctuate due to the above described correlations between the
different lumps like
\begin{equation}|\delta\rho|\approx \rho^{1/2}\end{equation}
The reason why we expect this kind of fluctuations to be more coherent
than the permanent ground fluctuations, due to almost randomly
changing bond-orientations among the elementary nodes, is the following. In
each lump or local cluster a certain fraction of the nodes/bonds is
slaved by the collective mode and behaves relatively coherently. By
the same token, \tit{interbonds} between nodes, belonging to these
respective sets in the various lumps, are also expected to change
their orientations more coherently than bonds, not belonging to these
particular sets. The same applies then to the blobs of charge,
exchanged via these particular bundles of interbonds connecting the
different lumps or local clusters. On the other
side, there are a great number of other lumps, our particular lump is connected
with via such bundles of interbonds. That is, the incoming or outgoing
blobs of charge have different phase relations as the local states in
the distant lumps or local clusters, our given lump is conneted with,
are different. We hence expect these fluctuations neither to be
completely correlated nor uncorrelated. We therefore arrive at the
following conclusion
\begin{conclusion} The charge fluctuations in a given lump, arising
  from the above described mechanism, are on the one side expected to be more
  coherent than the almost patternless groundfluctuations. On the
  other side, they should be sufficiently random on scale $L_{qm}$ to
  justify the above standard fluctuation formula.
We answer our first question by claiming that the quantum potential
arises just from this particular kind of non-local information
exchange between distant lumps.
\end{conclusion}  

Given that there exists such a type of fluctuation, with
$|\delta\rho|\approx\rho^{1/2}$, the next question is, how does this
fluctuation pattern effect a change of local frequency
\begin{equation}\delta(-\partial_tS(x,t))=-(\hbar^2/2m)\cdot\Delta(\rho^{1/2})\cdot\rho^{-1/2}\end{equation}
One may at first make up ones mind about how a mechanism can change at
all the local frequency with the corresponding amplitude, $Q_0+\varphi(x,t)$
being  more or less kept fixed. The period of the ground cycle is of order
$t_{pl}$. This time interval consists of a consecutive number of
\tit{clock-time intervals}, $\tau$,
\begin{equation}t_{pl}=N\cdot\tau\end{equation}
In each clock-time interval, $\tau$, the charge in a given lump, $P$,
is changed by the amount
\begin{equation}\Delta
  Q(t,P)=q(N_{in}(t;P)-N_{out}(t;P))=q(\sum_{in}J_{ik}(t;P)-\sum_{out}J_{ik}(t;P))\end{equation}
(cf. section 3.2). If this change per clock-time interval is increased
locally by some dynamical mechanism which results in a local change of
the network state in and around the lump under discussion, more
specifically, bigger positive $\Delta Q$ in the ascending part of the
cycle, bigger negative jumps in the descending part, with $Q_{max}$
kept essentially fixed by some other stabilizing mechanism, the local frequency
will increase and vice versa, since the necessary $Q_0$ is filled up
in a shorter or longer clock-time interval. In this sense one may envisage how
amplitude and frequency can vary more or less independently.

The quantum potential, and by the same token, its effect on the local
frequency, would vanish locally if $\Delta\rho^{1/2}=0$. On the other
side, we still expect the local amplitude to fluctuate on average by
the amount $\rho^{1/2}$. A locally constant $\rho$ means, according to
our interpretation, that the neigboring grains experience the same
amount of avergage fluctuation. We have now to remember that both the
ground frequency, $\omega_{pl}$ or the modulated frequency,
$\omega_{pl}+\partial_t S(x,t)$, and the corresponding amplitudes are
considered to be emergent quantities, being created by an autonomous
process of self-organisation within the network $QX/ST$. In other
words, these particular values within or around some lump or local
cluster, $C$, are the result of the local network state as a
whole. If this local environment is changed, we have to expect the same
for the local values of these collective quantities.

If $\Delta\rho^{1/2}\lessgtr 0$ around a given lump, the charge
fluctuation within the lump is greater/smaller on avergage compared to
the surrounding lumps (which can be inferred from Gauss-law).
According to our primordial network laws discussed in section 3.1 and
its implications on the level of the web of lumps, described in
section 3.2, higher charge fluctuation in a lump means a higher level
of reorientation of so-called \tit{interbonds} during a cycle of the
collective undulation.
\begin{obdef}
\begin{equation}\Delta\rho^{1/2}\lessgtr 0\end{equation}
leads to a higher/lower bond-fluctuation rate in the lump or the local
cluster under discussion compared to the neighborhood. We call this
bond-fluctuation rate the {\em bond-volatility}. It is another example
of an emergent collective quantity.
\end{obdef}
We conjecture now that such a higher/lower \tit{bond-volatility}
(relative to the surrounding lumps or local clusters) will enhance or
diminish the hight of the jumps we were talking about above and
thus increase or reduce the local frequency. We frankly agree that
this is, so far, only a qualitative analysis, but we know of related
effects in other fields of physics. See e.g. \cite{resonance} for a
\tit{stochastic triggering} or \tit{enhancement} of various
\tit{resonance phenomena}. We expect a similar mechanism to be at work
in our complex dynamical system.

A last point to mention is the physical meaning of the
``normalisation'' of the term by $\rho^{1/2}$ in the denominator. Such
small deviations from the huge vacuum values, we are talking about,
are so-called ``\tit{weak-field phenomena}''. We therefore expect that
there exists a \tit{linear} relation between the fluctuation of local
charge, $\rho^{1/2}$, and the number of nodes, being involved in this
fluctuation. One would hence get the change of frequency by following
the charge variation at a generic node over one cycle. One has hence
to divide by the amount of total charge fluctuation in the lump to get
the change in frequency. This explains the occurrence of the
denominator in the quantum potential.  The proportionality of the
respective quantities is, on the other side, encoded in the prefactor,
$\hbar^2/2m$.
\section{A Brief Commentary on Several-Particle Systems, the
  Transition towards Macroscopic Systems and State Vector Reduction}
In the preceding sections we have mainly discussed the
\tit{one-particle quantum theory}. We have omitted so far
\tit{several-} and \tit{many-body systems}, the transition to the
\tit{macroscopic regime} and the notorious and highly facetted quantum
mechanical \tit{measurement problem}. One reason for this restriction
was to keep the paper within reasonable length, since some of the
above mentioned topics have a long and venerable history of their own
and need a separate treatment. We want in the following to only
briefly indicate how we plan to procede in future work,in order to
cope with these problems.

In a first step we have to discuss the necessary changes which occur
in connection with several quantum objects, roaming through our
network. As to this set of questions we made already some preliminary
remarks in \ref{several}. Note also the critical attitude of
Heisenberg expressed in the utterance preceding our own conjecture
about the so-called \tit{problem of polydimensions}. What we try to
accomplish below is exactly to supply an interpretation \tit{of the
  results from polydimensions in three dimensions}, the possibility
which Schroedinger had in his mind and which Heisenberg considered to
be impossible. It becomes however apparent that the situation is much
more subtle than Schroedinger probably expected and that this task
cannot be accomplished in a naive sense (which Heisenberg rightly
criticized). It again turns out, that our concept of the two storeys
of space-time or the vacuum is crucial for solving this puzzle, a
concept, which Schroedinger did not yet have at his disposal!

Before going into the more technical details, we want to scrutinize
the above cited critical dictum of Heisenberg. Is it really impossible
to envisage the quantum mechanical several-particle situation in the
ordinary 3-dimensional cordinate space we are living in? Let us take a
classical $N$-particle system with $N$ sufficiently large and describe
it within the framework of classical statistical mechanics. It has, on
the one hand, several features which are similar to the quantum
mechanical wave-function representation, as both systems have many
elementary degrees of freedom. On the other hand, it has the advantage
that it can be completely understood!

Again, the so-called \tit{one-particle distribution function},
$\rho_1(x)$ is both a local probability and (as a \tit{particle
  density}) a local classical observable, that is, it has an immediate
interpretation in 3-dimensional coordinate space. The meaning of the
\tit{pair distribution function} or (modulo appropriate
normalisation) the \tit{two-particle correlation}, $\rho_2(x_1,x_2)$
is a little bit more abstract. In some loose sense it may be compared
with the $\rho(x_1,x_2)$ of two-particle Schroedinger theory. On the
one side, it figures as a function over abstract $\R^3\times\R^3$. On
the other hand, it describes a concrete feature of our \tit{compound
  system} as a whole, that is, a global property of our aggregate of
$N$ particles, living in the ordinary 3-dimensional coordinate
space. More specifically, it encodes the mutual influence of
the relative positions of the elementary constituents of our system on
each other. To put it succinctly: While formally being a function over
$\R^3\times\R^3$, it nevertheless encodes a concrete property of our
system living in three dimensions.

We now come to quantum theory. In the situation of a single quantum
excitation it was not necessary to go into the possible details of the
fine-structure of e.g. the charge modulations within or among the
lumps or local clusters, as only the total charge modulation,
$\delta Q_0(P,t)=a(P,t)$, entered in the coarse-grained equations of
low-energy quantum theory. In the several-particle theory (we discuss
in the following for convenience only the two-particel case)
$(\rho,S)$ depend now on two coordinates (see equations
(\ref{one}),(\ref{two}) ), in other words, they are now of an openly
\tit{non-local} nature in contrast to the one-particle case, where the
non-locality is more hidden, as has been described above.  

We may now envisage that we have in general two (or several)
\tit{entangled} excitation patterns, consisting of two
more or less distinguishable modulations of the high-frequency
ground-wave. These modulations are, as was already explained above,
\tit{weak-field excitations} on the comparatively large scale
$[L_{qm}]\gg [L_{pl}]$. As only a small fraction of the elementary
charges or nodes is involved in these weak deformation patterns, there
is ample space in the lumps (or local clusters) for approximately
individual pulse patterns to coexist and interact or influence each
other only weakly. They are however \tit{entangled} in general unless
the two-particle state is a \tit{product-state}, that is
\begin{equation}\phi(x_1,x_2)\neq
  \phi_1(x_1)\cdot\phi_2(x_2)\end{equation}

We now try to understand the mysterious phenomenon of entanglement.
\begin{conj}[Entanglement]What is called {\em entanglement} in
  ordinary quantum theory, is the {\em non-local} interaction
  (respectively, exchange of elementary charges) among the different
  excitation modes (describing the different quantum objects) via the
  bundles of {\em interbonds} connecting the different lumps or local
  clusters. That is, as in the above example of classical statistical
  mechanics, $\rho(x_1,x_2),S(x_1,x_2)$ describe no longer completely
  local properties of the system but express a complex non-local
  entanglement-pattern, permeating through our complicated {\em
    two-storey substrate}, $QX/ST$.
\end{conj} 
\begin{equation}\rho_1(x_1):=\int\rho(x_1,x_2)d^3x_2\end{equation}
collects e.g. all the different contributions, resulting both from
the entanglement with the components of the second excitation in the
other lumps and coming from the first excitation itself (as in the
one-particle situation). In the particular case of a \tit{free}
Hamiltonian and a product initial-state both excitations move
independently of each other and remain non-entangled, that is
\begin{equation}\rho(x_1,x_2)=\rho_1(x_1)\cdot\rho_2(x_2)\end{equation}
In this case each particle receives \tit{non-local information} only
from its own distant components in the other lumps as described
above.\\[0.3cm]
Remark: We assume, for reasons of simplicity (as this section is only
of a cursory and preparatory nature), that the particles are
distinguishable.\\[0.3cm]
Note however that a \tit{non-product state}, $\rho(x_1,\ldots,x_n)$,
cannot be uniquely decomposed into  a product state and a contribution
encoding the \tit{entanglement} with the other particle-excitations in the
distant lumps. In any case for fixed $(x_1,\ldots,x_n)$, in the
entangled situation, a certain fraction of the elementary charges
belongs, so to say, to all the involved lumps at the same time as they
oscillate between them via the bundles of respective interbonds. These
and related phenomena have to be analyzed more carefully in future
work.
\begin{conclusion}The above cursory analysis shows, that ``{\em the
    results from polydimensions}'' can be understood in a satisfactory
  way in three dimensions under the proviso that we accept the {\em
    two-storey structure} of space-time or the physical vacuum.
\end{conclusion}

In a next step one can envisage what happens if the number of these
coexisting excitations become too large in a given volume of space. In
a \tit{many-body wave-function}
\begin{equation}\psi(x_1,\ldots,x_N)\quad N\gg 1\end{equation}
each of the approximately individual excitation modes has to be
entangled \tit{non-locally} with the other modes and/or with its own
components in the distant lumps. If $N$ exceeds a certain critical
range of values, a complete quantum-entanglement may not longer be
possible. The \tit{weak-field picture} may begin to break down as all the
available information channels (that is, the interbonds among the
different lumps) are occupied or overcrowded. The wave-function will
start to decay and goes over into a (partial mixture), that is, an
incompletely entangled state.

A last point to mention is the infamous \tit{measurement problem}. The
seemingly instantaneous \tit{state-reduction} by a measurement
interference (in fact the contact with a peculiarly tuned macroscopic
apparatus), while being on the surface of a quasi-local nature, will
nevertheless spread its corresponding (decoherence-)information all
over the microsystem almost instantaneously via the network of
existing interbonds. It is thus yet another manifestation of the
peculiar non-local character of the vacuum, described in the preceding
sections.\\[0.3cm]
Remark: An, in our view, quite up to date discussion of some of the
impending problems may be found in section 2 of \cite{Haag}.

\end{document}